\documentclass[apj,twocolumn,twocolappendix,floatfix]{openjournal}

\DeclareRobustCommand{\VAN}[3]{#2}
\let\VANthebibliography\thebibliography
\def\thebibliography{\DeclareRobustCommand{\VAN}[3]{##3}\VANthebibliography}

\usepackage{amsmath}	
\usepackage{amssymb}	
\usepackage{booktabs}
\usepackage{lipsum}
\usepackage[T1]{fontenc}
\usepackage{graphicx}	
\usepackage[breaklinks,colorlinks,citecolor=blue,urlcolor=blue,linkcolor=blue]{hyperref}
\usepackage{cleveref}
\usepackage{multirow}
\usepackage{newtxtext,newtxmath}
\usepackage{orcidlink}
\usepackage{pifont}
\usepackage[nobottomtitles]{titlesec}
\usepackage{natbib}
\usepackage{xparse}
\usepackage{ifthen}
\titleformat{\section}{\filcenter\MakeUppercase}{\thesection.}{0.5em}{}
\usepackage{wrapfig}
\usepackage{savesym}
\savesymbol{tablenum}
\usepackage{textcomp}
\usepackage{siunitx}
\restoresymbol{SIX}{tablenum}
\usepackage{needspace}
\usepackage{overpic}
\usepackage{xcolor}
\usepackage{pgffor}
\usepackage{float}
\usepackage{footmisc}

\newcommand{\Gaia}{\textit{Gaia}}
\newcommand{\TESS}{\textit{TESS}}
\newcommand{\WISE}{\textit{WISE}}
\newcommand{\GALEX}{\textit{GALEX}}
\newcommand{\ipd}{\texttt{ipd\_frac\_multi\_peak}}

\makeatletter
\renewcommand{\fnum@figure}[1]{Fig.\ \thefigure:}
\makeatother

\newcommand{\xpos}{1}
\newcommand{\ypos}{1}
\newcommand{\putbox}[3]{\put(#1,#2){\Large\textbf{(#3)}}}
\newcommand{\panelfigwidth}{0.31\linewidth}

\begin{document}
\title[Variable Stars]{A systematic search for big dippers in ASAS-SN}


\author{\vspace{-1.3cm}B. JoHantgen\,$^{1}$,
D. M. Rowan\,$^{1,2}$,
R. For{\'e}s-Toribio\,$^{1,2}$,
C. S. Kochanek\,$^{1,2}$,
K. Z. Stanek\,$^{1,2}$,\\
B. J. Shappee\,$^{3}$,
Subo Dong\,$^{4,5,6}$,
J. L. Prieto\,$^{7,8}$, and
Todd A. Thompson\,$^{1,2,9}$}

\affiliation{$^{1}$Department of Astronomy, The Ohio State University, 140 W. 18th Avenue, Columbus, OH 43210, USA \\ 
$^{2}$Center for Cosmology and Astroparticle Physics, The Ohio State University, 191 W. Woodruff Avenue, Columbus, OH 43210, USA \\
$^{3}$Institute for Astronomy, University of Hawaii, 2680 Woodlawn Drive, Honolulu, HI 96822, USA \\
$^{4}$Department of Astronomy, School of Physics, Peking University, 5 Yiheyuan Road, Haidian District, Beijing 100871, People's Republic of China \\
$^{5}$The Kavli Institute for Astronomy and Astrophysics, Peking University, 5 Yiheyuan Road, Haidian District, Beijing 100871, People's Republic of China \\
$^{6}$National Astronomical Observatories, Chinese Academy of Science, Beijing 100101, People's Republic of China \\
$^{7}$Instituto de Estudios Astrf{\'i}sicos, Facultad de Ingenier{\'i}a y Ciencias, Universidad Diego Portales, Avenida Ejercito Libertador 441, Santiago, Chile \\
$^{8}$Millennium Institute of Astrophysics MAS, Nuncio Monse{\~n}or Sotero Sanz 100, Off. 104, Providencia, Santiago, Chile \\
$^{9}$Department of Physics, Ohio State University, 191 W. Woodruff Ave., Columbus, OH 43210, USA \\}

\begin{abstract} 
Dipper stars are extrinsically variable stars with deep dimming events due to extended, often dusty, structures produced by a wide range of mechanisms such as collisions, protoplanetary evolution or stellar winds. ASAS-SN has discovered 12 dipper-like objects as part of its normal operations. Here we systematically search the $\sim 5.1$ million ASAS-SN targets with $13<g<14$~mag for dippers with $\Delta g\ge0.3$~mag to identify 4 new candidates. We also discover 15 long-period eclipsing binary candidates. We characterized the 19 new and 12 previously discovered objects using the ASAS-SN light curves and archival multi-wavelength data. We divide them into three categories: long-period eclipsing binaries with a single eclipse (13 total), long-period eclipsing binaries with multiple eclipses (7 total) and dipper stars with dust or disk occultations (11 total).
\end{abstract}


\maketitle

\section{Introduction} \label{sec:introduction}

Variable stars offer a valuable opportunity to study the structure and evolution of stars and stellar populations. For example, intrinsic variables, like RR Lyrae and Cepheid pulsators, are used to study the internal structure of the stars \citep{Marconi_2015, Caputo_2000}. Extrinsic variables, like rotational variables and eclipsing binaries, are stars that are variable due to occultations from other objects or features on the star. Eclipsing binaries are important tools for measuring the masses and radii of stars \citep{Andersen_1991, Torres_2010, Rowan_2025}, and rotational variables can inform models of chromospheric activity and magnetic fields \citep{Cheng_1986}, and can be used in gyrochronology to determine the age of stellar systems \citep{Barnes_2007}.

Transient surveys generally look for rapidly brightening sources such as supernovae, and variability surveys look for continuously variable sources such as eclipsing binaries, pulsators, and rotational variables. There are few systematic searches for "occasionally" dimming sources. Young stellar objects (YSOs) like T Tauri stars \citep{Bredall_2020} and R Coronae Borealis stars \citep[R Cor Bor stars,][]{Crawford_2025} can show dramatic fading events; however, these events are recurrent. 

One class of occasionally dimming sources show long, deep, dimming events caused by disks. These objects are often discussed in the context of $\epsilon$ Aurigae \citep{Carroll_1991} and KH 15D \citep{Kearns_1998}. $\epsilon$ Aurigae has a period of 27.1 years, with dips that reach a depth of $\Delta V \sim 0.75$~mag and last for about 2 years due to an occultaion by a companion B star with a disk \citep{Stefanik_2010}. KH 15D has a period of 48 days, but has variable eclipses because of the inclination of its circumbinary disk \citep{Winn_2006, Poon_2021}. However, this classification seems to be insufficient to explain the wealth of configurations that can lead to such long and deep events \citep[see ][]{Fores-Toribio_2025}. OGLE-LMC-ECL-11893 is another object that features eclipses caused by a disk; however, the eclipses have the same shape over many periods, which is not typical for disk eclipsing systems \citep{Dong_2014}. Boyajian's star \citep{Boyajian_2016, Simon_2018} and Zachy's star \citep{Way_2019} are more examples of objects that have long and deep dimming events. The Via Lactea (VVV) Survey \citep{Minniti_2010} carried out a search for objects with large amplitude dimming and continuous variability in \cite{Lucas_2024}. Their search discovered 40 YSOs and 21 dipping giant star candidates. These objects are further categorized and characterized in \cite{Guo_2024}.

The All-Sky Automated Survey for SuperNovae \citep[ASAS-SN, ][]{Shappee2014, Kochanek2017, Hart2023} is a transient survey that can also be used to search for long period EBs and dipper stars. It has a baseline of roughly 10 years and, in its current configuration of 20 telescopes on 5 mounts across 4 sites (Chile, Hawaii, South Africa, and Texas), it achieves a sub-daily cadence across the visible sky in good conditions. ASAS-SN has discovered several "occasional" dimming events. Examples include ASASSN-21co, an eclipsing binary with a period of 11.9 years \citep{Rowan2021}, and ASASSN-23ht, an object with one deep dimming event in the ASAS-SN photometry \citep{Miller_2023}. A second class, which we refer to as "big dipper stars", show deep ($\Delta g\ge0.3$~mag) dimming events that can last tens to hundreds of days. These objects are distinct from the dippers defined by \cite{Ansdell_2016}. In the context of this work, dippers are objects with non-recurrent, non-symmetric dimming events that can be long-lasting and have a wide variety of features. These events are likely caused by occultations by extended dusty structures that are produced in a variety of different mechanisms. Examples include ASASSN-24fw, an object with a deep dimming event caused by a cloud of dust \citep{Fores-Toribio_2025, Zakamska_2025}, and ASASSN-21qj, a star with a deep, long-lasting event that is thought to be caused by dust formed in a collision of objects orbiting the star \citep{Kenworthy2023, Marshall_2023}. 

Here we carry out a systematic search in ASAS-SN for stars that have "occasional" dimming events. Section \S\ref{sec:methods} describes our search strategy and the types of false positives. We characterize the serendipitously found and detected targets in Section \S\ref{sec:discussion of each target} using additional broadband photometry, large spectroscopic surveys, and X-ray catalogs. Finally, Section \S\ref{sec:Conclusions from the targets} summarizes how this search can be expanded using other photometric surveys and machine learning methods.  

\section{Methods} \label{sec:methods}

\begin{figure*}
    \centering
    \begin{tabular}{ccc}
        \begin{overpic}[width=\panelfigwidth]{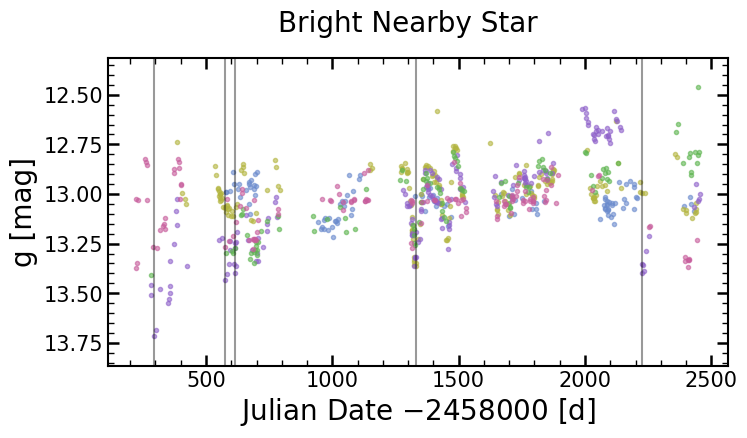}
            \putbox{\xpos}{\ypos}{a}
        \end{overpic} &
        \begin{overpic}[width=\panelfigwidth]{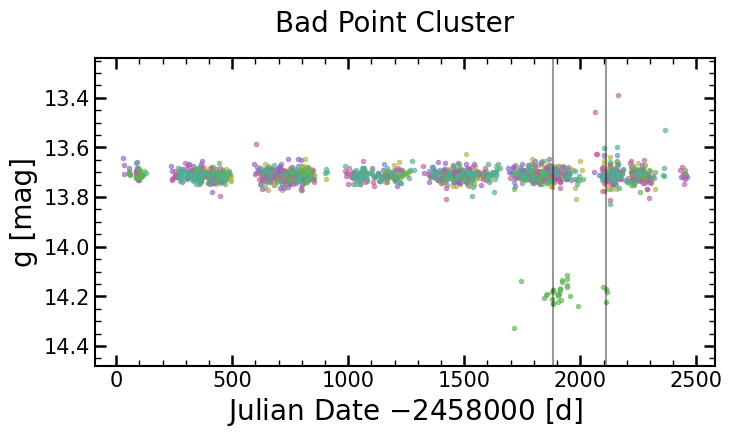}
            \putbox{\xpos}{\ypos}{b}
        \end{overpic} &
        \begin{overpic}[width=\panelfigwidth]{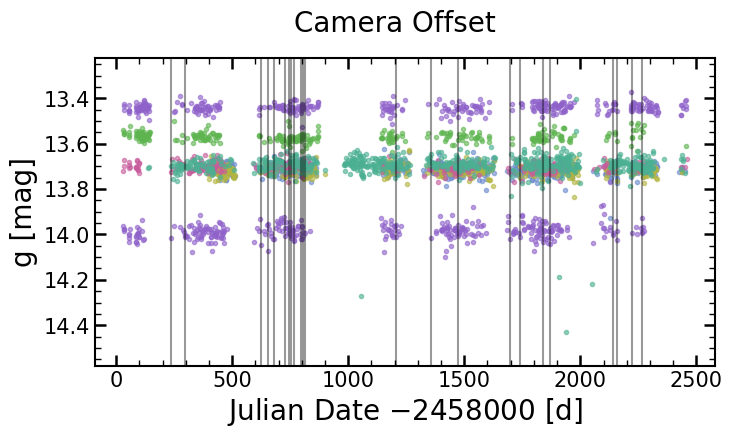}
            \putbox{\xpos}{\ypos}{c}
        \end{overpic} \\
        \begin{overpic}[width=\panelfigwidth]{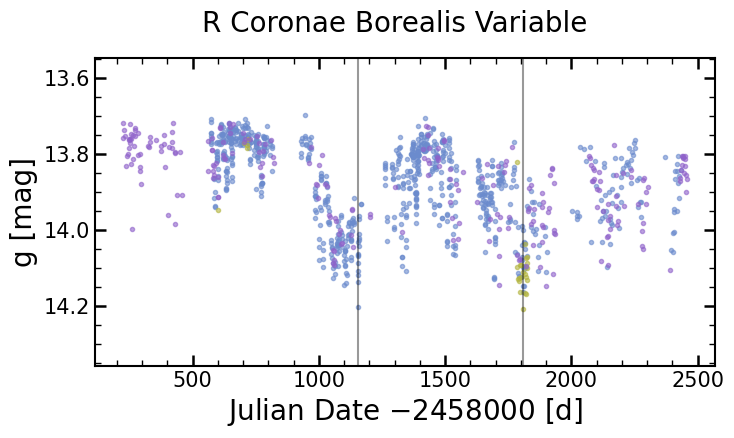}
            \putbox{\xpos}{\ypos}{d}
        \end{overpic} &
        \begin{overpic}[width=\panelfigwidth]{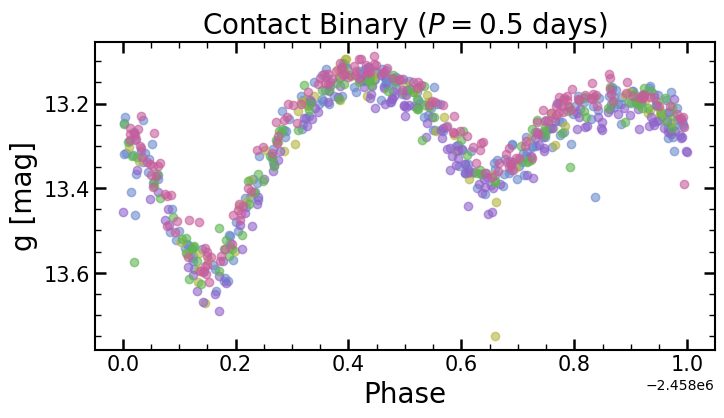}
            \putbox{\xpos}{\ypos}{e}
        \end{overpic} &
        \begin{overpic}[width=\panelfigwidth]{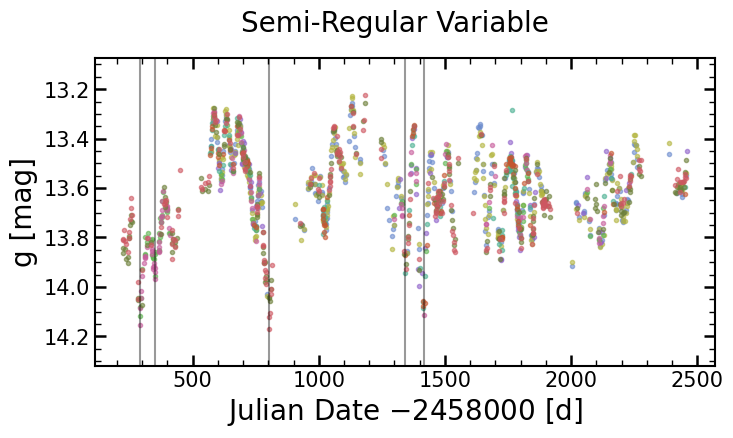}
            \putbox{\xpos}{\ypos}{f}
        \end{overpic} \\
        \multicolumn{3}{c}{
            \begin{tabular}{c}
                \begin{overpic}[width=0.35\textwidth]{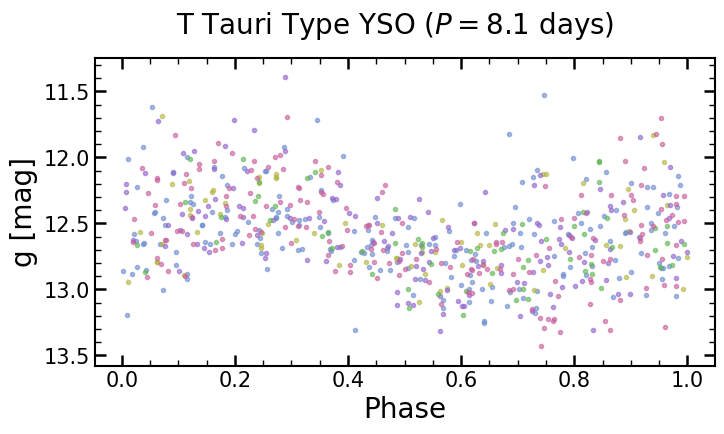}
                    \putbox{\xpos}{\ypos}{g}
                \end{overpic}
            \end{tabular}
        }
    \end{tabular}
    \caption{Examples of false positives that were rejected during the visual inspection of possible candidates and objects rejected by our search labeled by their rejection reason. Some of the rejections are due to other types of variability, some are caused by a few bad points, and some effects are environmental. The vertical lines correspond to the dimming events detected by our pipeline. The different colors correspond to different cameras. In panels (e) and (g), the light curves are phase-folded using the period from a Lomb-Scargle periodogram
    \citep{Lomb_1976,Scargle_1982}.}
    \label{fig:False_Positives}
\end{figure*}

We obtained the ASAS-SN light curves of the 5.1 million sources with $13<g<14$~mag using SkyPatrol v2 \citep{Hart2023}. This magnitude range was selected because it provided a significant sample of unsaturated, fairly high signal-to-noise ratio light curves for an initial search. We use the $g$-band light curves to search for candidates, but include the $V$-band photometry to characterize some of the candidates. We improve the calibrations between the ASAS-SN cameras using a damped random walk Gaussian process for interpolation to optimize the camera offsets \citep[see, e.g.,][]{Kozlowski_2010}. We then designed a pipeline to find all objects that had been serendipitously discovered by ASAS-SN. We searched the light curves for variations with $\Delta g\ge0.3$~mag. A detected dip was defined as at least one point that satisfied the depth condition. We then removed all stars with light curve dispersions $\sigma > 0.15$~mag. This removes many common variables with high amplitude, \mbox{(quasi-)}continuous variability (e.g., shorter period EBs, YSOs, R Cor Bor stars). We initially allowed events to be in only one camera. As we started visual inspection, it was clear that we could reduce the false positive rate by requiring dips in multiple cameras with temporal separations of less than one day. With this change, we had $\sim11{,}000$ candidates, of which $\sim1{,}400$ were known variables. These were all visually inspected.

Figure \ref{fig:False_Positives} shows examples of false positives rejected during visual inspection. Systematic false positives can be seen in panels (a) through (c) and astrophysical false positives can be seen in panels (d) through (g). Some show continuous variability, but not "occasional" events. Figure \ref{fig:False_Positives}a shows the light curve of a star too close to a bright star. In many cases this is due to the bleed traits of quite distant saturated stars. These make up $\sim30\%$ of our false positives. In Figure \ref{fig:False_Positives}b there is a group of problematic data from a specific camera that triggered its selection. This is the problem leading to the requirement of detections in multiple cameras. Roughly $\sim20\%$ of our false positives are due to this issue. Figure \ref{fig:False_Positives}c shows an example of a failure mode that only occurs near the poles. A source almost always appears in only one field for a given camera, so the light curve intercalibration procedure assumed a single offset for each camera. However, close to the equatorial poles, field rotations can allow a source to appear in two fields for the same camera, leading to problems if the fields need different offsets. This error caused $\sim28\%$ of our false positives. Panels (d) through (g) of Figure \ref{fig:False_Positives} are variable stars. Panel (d) is an R Coronae Borealis variable and panels (e) through (g) are more common contaminates (contact binaries, semi-regular variables, and YSOs, respectively). Known variables made up $\sim22\%$ of the false positives.

\begin{figure}[t!]
    \centering
    \includegraphics[width=\linewidth]{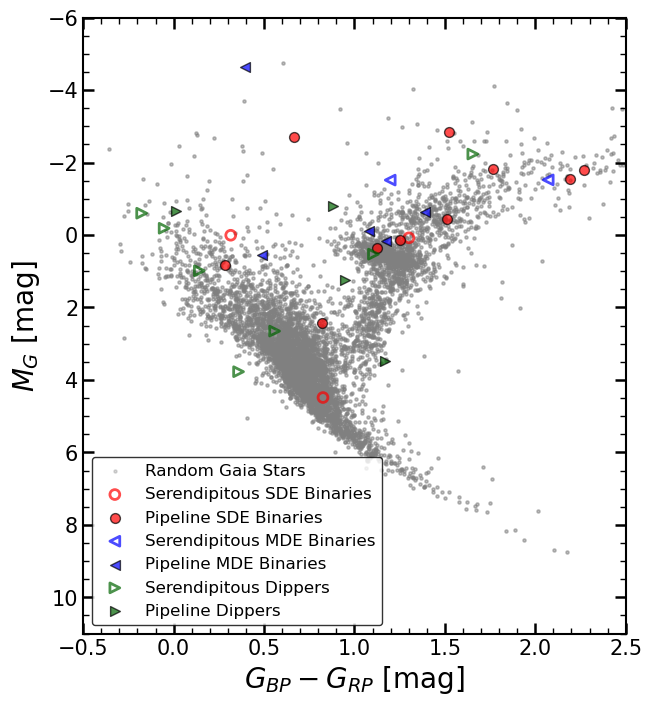}
    \caption{The \Gaia \space CMD with the known sources and the new sources. The targets on the CMD are sorted into the categories of single-dip eclipsing binaries (SDE binaries) candidates, multi-dip eclipsing binaries (MDE binaries) candidates, and dipper candidates. The grey background stars show the first $\sim100{,}000$ stars run through our pipeline.}
    \label{fig:cmd}
\end{figure}

In total, we found 19 new candidates and recover 11 of the 12 previously known systems. J114712$-$621037 was not recovered because its dipping event is shallower than $0.3$~mag and had a very long time scale ($\sim$1000 days). While we could lower the required magnitude change threshold of $\Delta g >0.3$, this would have significantly increased the number of targets that would need to be visually inspected. Both the previously known objects and the newly discovered objects are shown on a \Gaia \space color-magnitude diagram (CMD) in Figure \ref{fig:cmd} \citep{GaiaColab, GaiaDR3} and Table \ref{table:target_param} lists the sources, labeled by whether they were discovered earlier as a part of daily ASAS-SN transient search or using our pipeline, along with the \Gaia \space absolute magnitude and colors. We compute extinctions using the {\tt mwdust} 3-dimensional dust map \citep[][which is based on, \cite{Marshall_2006}, \cite{Green_2015}, and \cite{Drimmel_2003}]{Bovy_2016}, and use distances from \cite{Bailer-Jones_2021}. Table \ref{table:dip_param} lists the number of dips found, the maximum dip duration, the epoch of the most recent dip and the maximum dip depth in the $g$-band. The two tables divide the sources into three categories: probable EBs with only a single eclipse, probable EBs with multiple eclipses, and dipper stars. The light curves for each of the final candidates that had passed the visual inspection can be found in a supplemental folder accompanying this work. This folder includes zoomed in sections of each light curve corresponding to where the pipeline was triggered.

We checked the earlier $V$-band ASAS-SN data for each of our final candidates to look for additional dimming events. We also checked their ALLWISE/NEOWISE \citep{Wright2010, Mainzer_2011} light curves, extracted using tools from \cite{Hwang_2020}. One source, J183210$-$173432, lacked \WISE \space data due to the proximity of a bright star. We searched VizieR \citep{Ochsenbein_2000} for archival data on each source, in particular for reports of X-ray detections, infrared excesses, and prior variability. We examined LAMOST \citep{lamost} and APOGEE \citep{Eisenstein_2011} spectra when available to search for Balmer (or other) emission lines. We used \WISE, 2MASS \citep{Cutri03}, \GALEX \space \citep{Bianchi11}, and \Gaia{} \citep{Gaia23_gspc} photometry to examine the source spectral energy distributions (SEDs), in particular to look for mid-IR emission excesses relative to \cite{Castelli03} model stellar atmospheres from {\tt pystellibs}\footnote{\url{https://github.com/mfouesneau/pystellibs}} indicating the presence of circumstellar dust. Finally, for targets flagged as variable stars in other catalogs, we use a Lomb-Scargle periodogram \citep{Lomb_1976, Scargle_1982} to determine the period.

We used two \Gaia{} statistics to search for evidence of binarity. The \Gaia{} parameter RUWE (renormalized unit weight error) is related to the astrometric goodness of fit and a value higher than 1.4 is a good indicator of binarity \citep{GaiaColab, GaiaDR3}. The Gaia parameter {\tt rv\_amplitude\_robust} ($RV_{amp}$), is the peak-to-peak amplitude of the Gaia RV measurements after outlier rejection. It can be used to identify binaries without orbital solutions, with systems having $RV_{amp} > 20$ km/s very likely to be binaries \citep{GaiaColab, GaiaDR3}.


\section{Discussion of each target} \label{sec:discussion of each target}

We divide our candidates into three categories; single-dip eclipsing binary candidates, multi-dip eclipsing binary candidates, and big dippers. For each category, we first summarize the objects discovered in ASAS-SN. For systems in the first two classes, we can also crudely estimate the period with a Gaia measurement of $RV_{amp}$. In an equal mass binary, the observed peak-to-peak velocity amplitude of an edge-on, circular orbit is
\begin{equation}
    V_p= \frac{2\pi a}{P}= \left(\frac{2\pi  G M_T}{P}\right)^{1/3} = 213 M_{T\odot}^{1/3} P_d^{-1/3}~\hbox{km/s}
    \label{eqn:MPtoV}
\end{equation}
where $a$ is the semimajor axis and $M_T$ ($M_{T\odot}$) is the total system mass (in solar masses) and $P$ ($P_d$) is the period (in days). We observe $V_p \sin i$, where $i$ is the inclination, but for wide eclipsing binaries with $\sin i \simeq 1$, we can estimate a minimum system mass by comparing $V_p$ and $RV_{amp}$. Assuming the RVs are measured for the more massive star (i.e., the most luminous), $V_p$ can be reduced by increasing the mass ratio and ellipticity can increase it by at most $(1-e^2)^{-1/2}$. The amplitude can also be underestimated because the RV measurements do not fully sample orbits longer than the roughly 1000~day (2.8~years) span of the data in \Gaia{} DR3. Once the Gaia data are no longer sampling the full orbit, Eqn.~\ref{eqn:MPtoV} will begin to increasingly overestimate periods. Another issue is that once $RV_{amp}$ gets close to $\sim20$~km/s, much of it could be created by noise in the RV measurements, leading to Eqn.~\ref{eqn:MPtoV} to underestimate the period.

\subsection{Single-Dip Eclipsing Binaries} 

\begin{figure*}
    \centering
    \includegraphics[width=\linewidth]{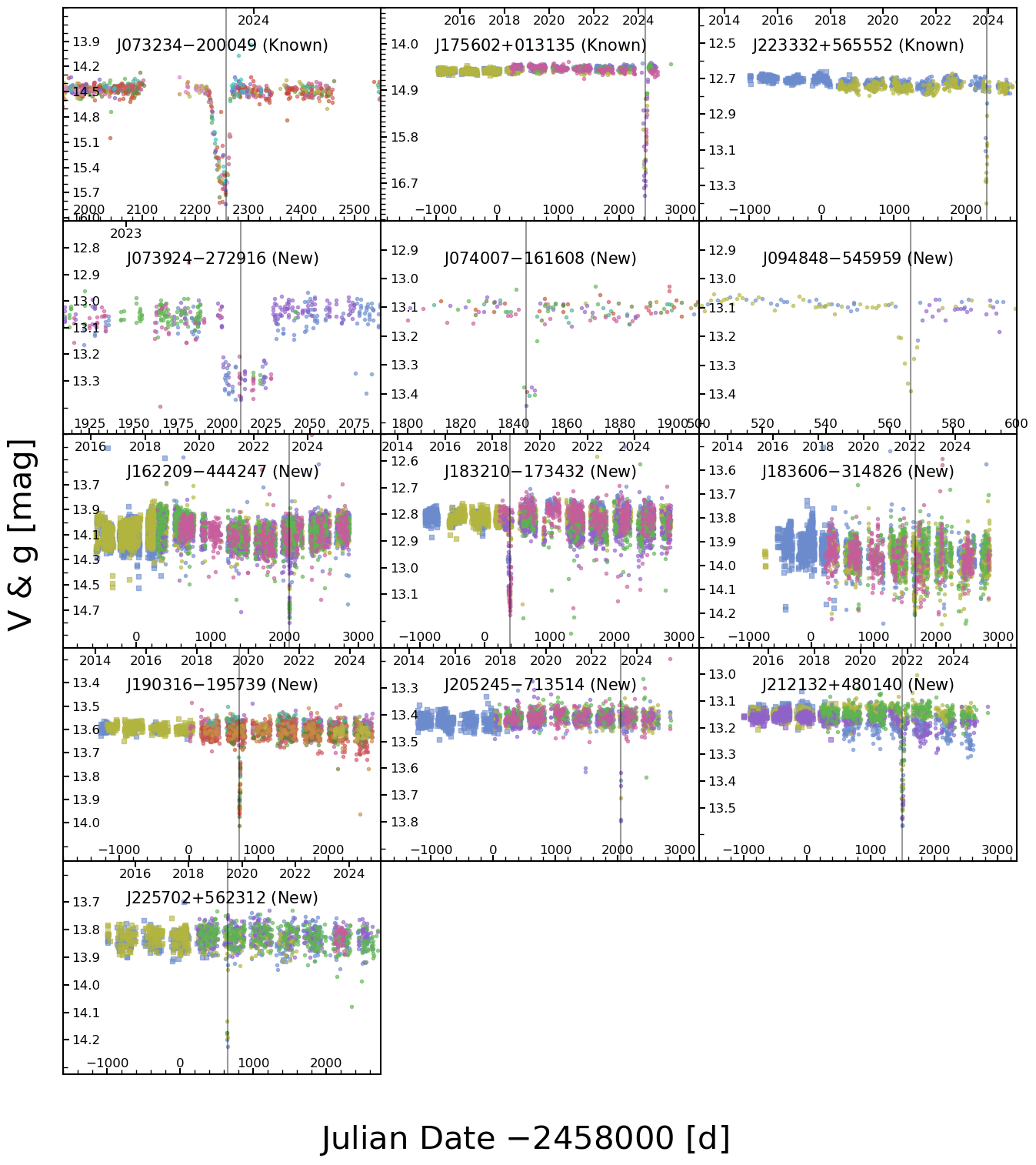}
    \caption{The single-dip eclipsing binaries discussed in Section 3.1. The serendipitously found targets are shown first, followed by the targets found in this search, both ordered by right ascension. The first known target shows the light curve zoomed in on the dip, while the rest of the known targets show the full ASAS-SN light curves. The first three new targets show the light curves, zoomed in on the dips while the remaining new targets show the full ASAS-SN light curves. The zoomed in cases are to better show the shape of the events. Each different color in the light curve represents a different ASAS-SN camera. The circles (squares) are the ASAS-SN $g$ ($V$) band data.}
    \label{fig:SDE_Binaries}
\end{figure*}

The single-dip eclipsing binaries are objects that have a single symmetric dip in their light curves. The shapes of the multi-camera eclipse light curve, such as having a symmetric ingress and egress, are used to visually confirm an eclipse. The symmetric shape of the dip implies that the occulting object has a regular shape, like a star, rather than an asymmetric or non-uniform dusty occulter. As only one dip is seen in the light curve, any other eclipses either occur in observational gaps, or the period of the binary is longer than the operational time of ASAS-SN ($\sim10$ years). The deepest of these single eclipses may be dippers, rather than EBs, but they are symmetric and have depths within the limits possible for EBs \citep[see ][]{Fores-Toribio_2025}. Deeper dips and longer periods become increasingly unlikely to be EBs. We folded the light curves as a function of period to find the minimum period where prior eclipses could all fit in seasonal gaps or prior to the start of the survey. They are presented in Table \ref{table:dip_param}.

J073234$-$200049 \citep[ASASSN-23ht, Figure \ref{fig:SDE_Binaries},][]{Miller_2023} is an upper main sequence star with a single dimming event that began on 09/29/2023 and lasted for about 114 days. It reached a depth of $\Delta g\simeq1.2$~mag. This eclipse depth and duration combination are possible for an EB, and the eclipse is close to being symmetric, but it could plausibly be a dipper star instead of an EB. The minimum period based on the light curve is $\ge8$ years.

J175602$+$013135 (ASASSN-24cf, ASAS-SN Transient Page\footnote{\label{fn:transients}\url{https://www.astronomy.ohio-state.edu/asassn/transients.html}}) is on the giant branch and has a single dimming event that began on 22/02/2024 and lasted for about 95 days with a maximum depth of $\Delta g\simeq2.3$~mag. This object is identified as a RS Canum Venaticorum (RS CVn) variable in \Gaia \space \citep{GaiaColab, GaiaDR3}. It is difficult for EBs to have this combination of depth and duration, but not impossible.

J223332$+$565552 (ASASSN-23ik, {ASAS-SN Transient Page}\footnotemark[\ref{fn:transients}]) is a lower main sequence star with a single dimming event that began 15/11/2023 and lasted for about 20 days with a maximum depth of $\Delta g\simeq0.7$~mag. It has been classified as a $\beta$ Persei-type (Algol) eclipsing system, however, there is no reported period \citep{vsx}. With an $RV_{amp}=22.29$~km/s, we can estimate a period of $P=873M_{T\odot}$~days, using Eqn.~\ref{eqn:MPtoV} and assuming an edge-on circular orbit. However, we see no other events in either the $g$ or $V$ band light curves with an estimated minimum period of roughly 1700 days. This minimum period is close to two times the estimate using Eqn.~\ref{eqn:MPtoV}, however, the $RV_{amp}$ is close to the threshold of $20$~km/s where underestimates of the period become more likely.

J073924$-$272916 had a single, $\Delta g\simeq0.3$~mag dimming event that began on 25/02/2023 and lasted for about 27 days. It is a red giant and the event is similar to the event seen in J181752$-$580749 below. ATLAS reports this target as a "dubious" binary with $P=2.01$ days \citep{Heinze18}. With an $RV_{amp}=23.91$~km/s, the period estimate using Eqn.~\ref{eqn:MPtoV} is $P=707M_{T\odot}$~days. Which is likely an underestimate. From folding the light curve we estimate a minimum period of $\sim1{,}300$~days.

J074007$-$161608 had a single, $\Delta g\simeq0.3$~mag dimming event on 21/09/2022 that lasted about 4 days. It is a luminous star in the Hertzsprung Gap. Its CMD position is uncertain because its \Gaia \space parallax uncertainty is large ($\varpi = 0.0742 \pm 0.0321$ mas), which also affects the extinction estimate. The events are similar to those in J181752$-$580749 and J223332$+$565552. We do not see any additional events. 

\begin{figure}
    \centering
    \begin{overpic}[width=\linewidth]{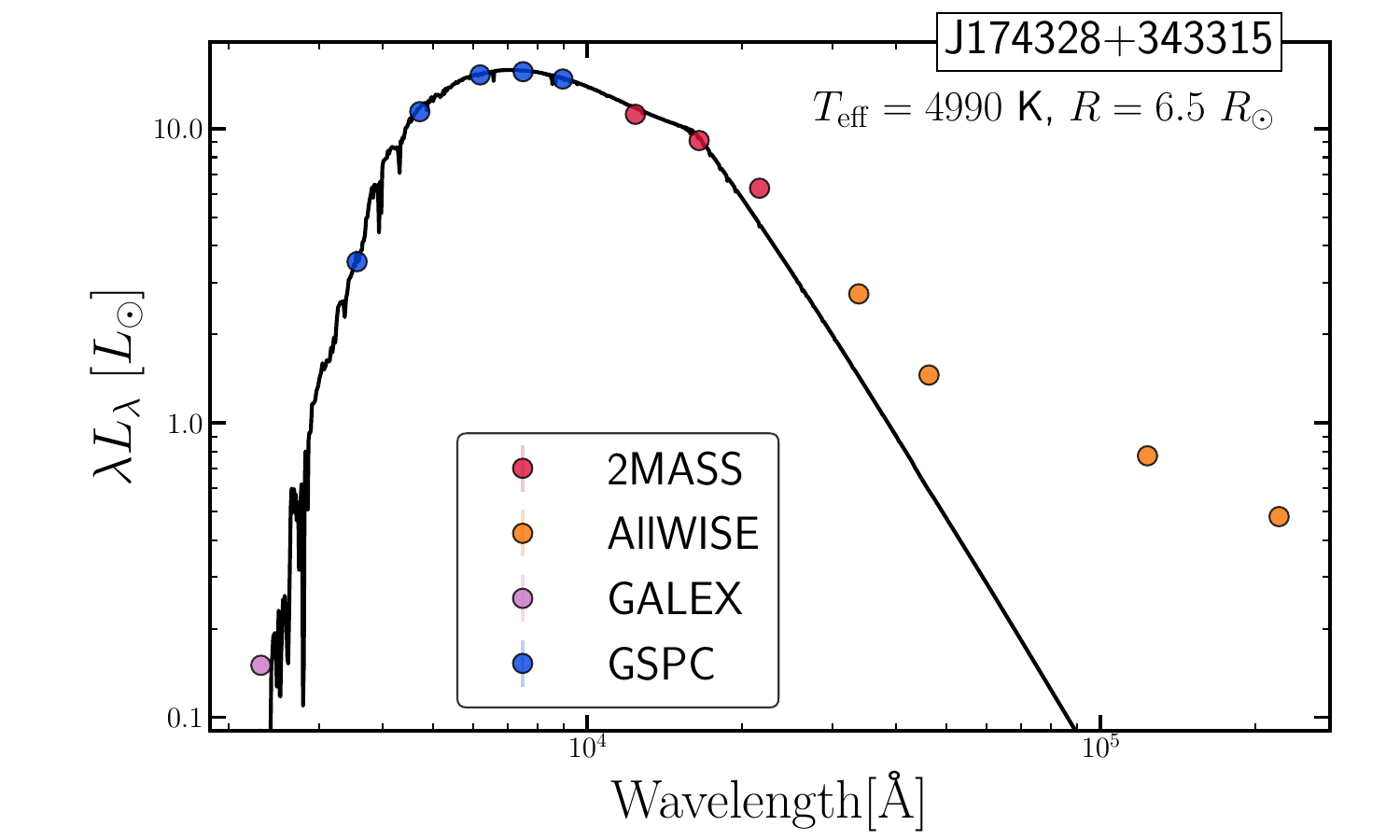}
        \putbox{\xpos}{\ypos}{a}
    \end{overpic} 
    \begin{overpic}[width=\linewidth]{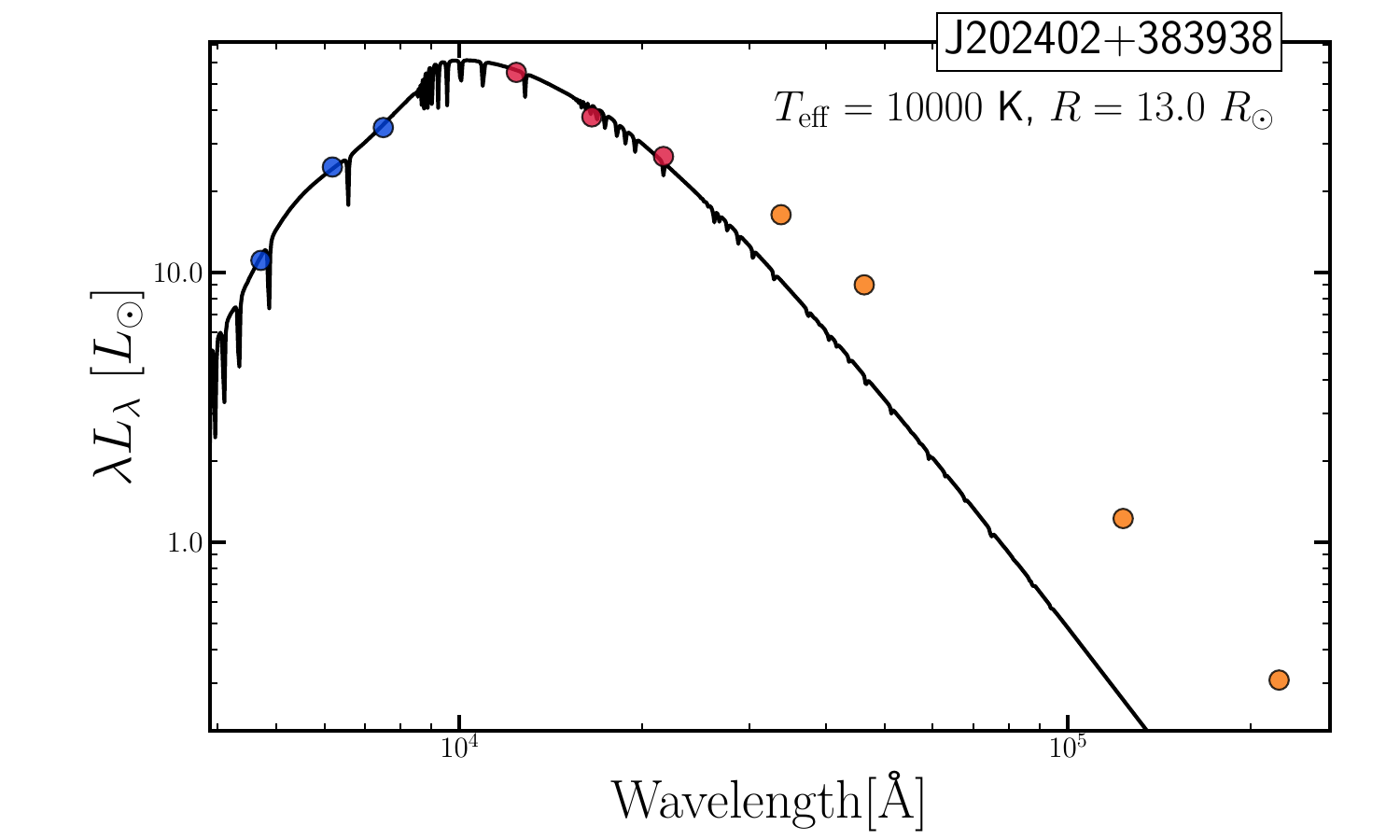}
        \putbox{\xpos}{\ypos}{b}
    \end{overpic} 
    \begin{overpic}[width=\linewidth]{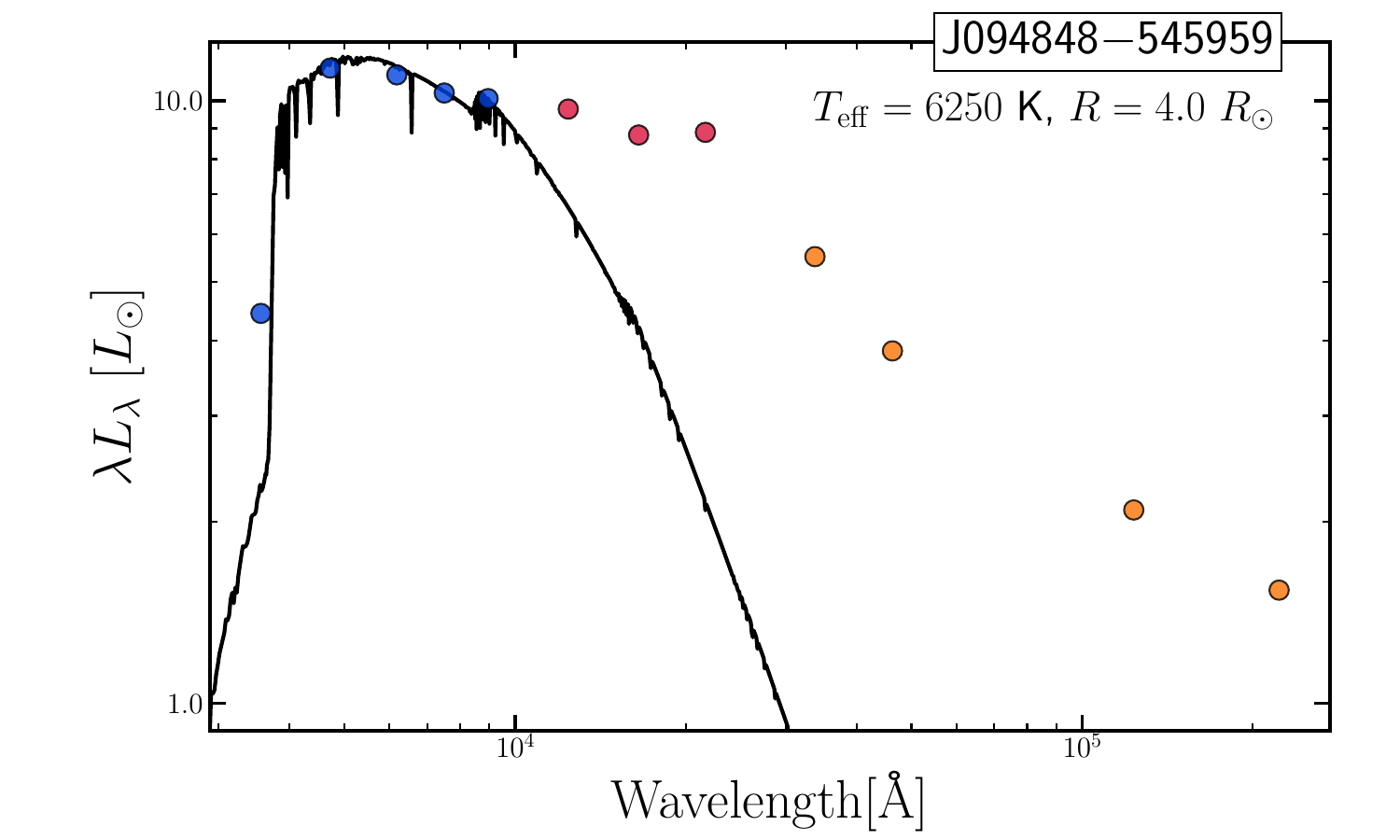}
        \putbox{\xpos}{\ypos}{c}
    \end{overpic}
    \begin{overpic}[width=\linewidth]{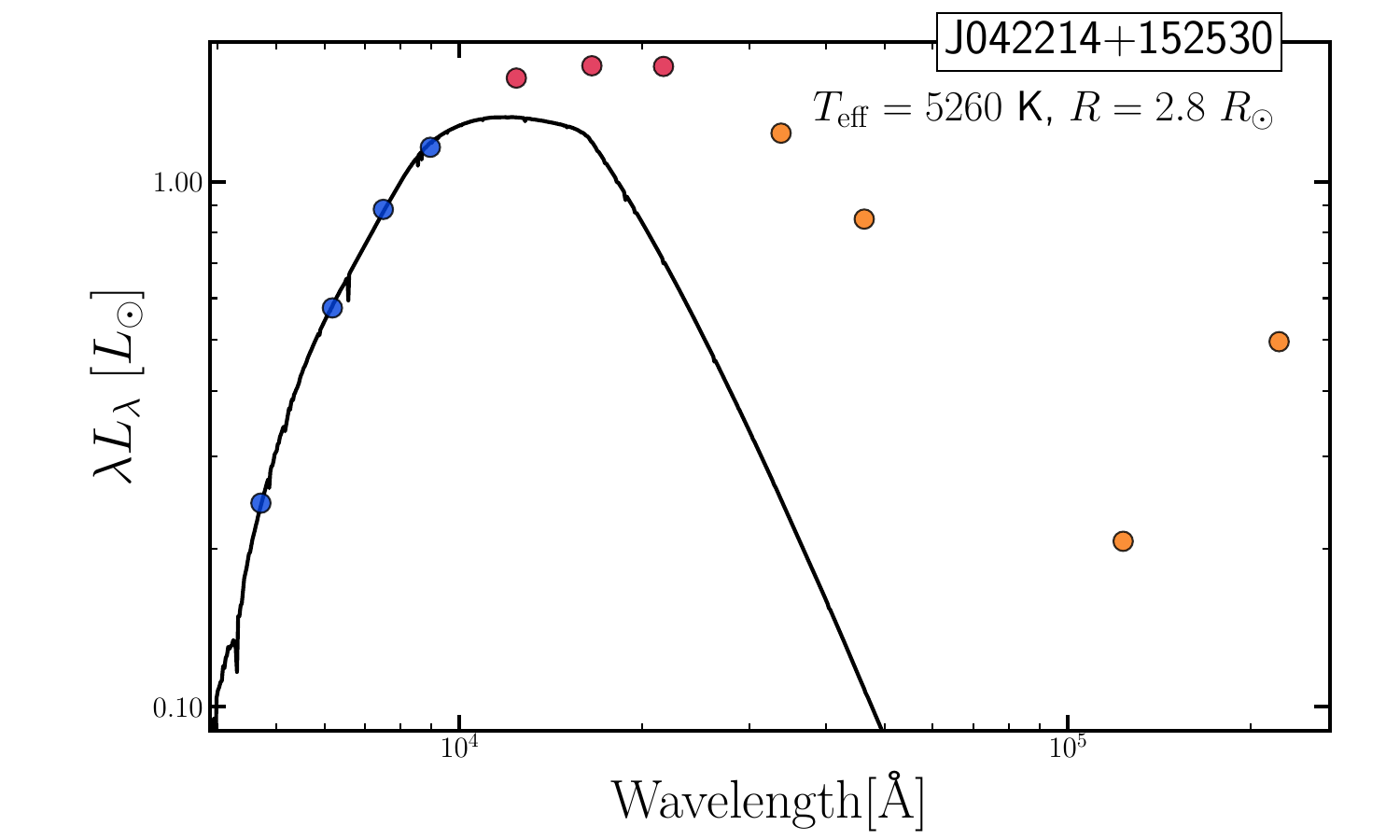}
        \putbox{\xpos}{\ypos}{d}
    \end{overpic} 
    \caption{Spectral energy distributions of the dippers that have significant IR excesses. The black lines are single star SED models fit to the optical part of the SED. The effective temperature and radius of the model is reported in the upper right corner of each panel.}
    \label{fig:SED_dippers}
\end{figure}

J094848$-$545959 had a single, $\Delta g\simeq0.3$~mag dimming event on 20/03/2019 that lasted for about 8 days. We do not see any additional events; the features near $\rm{JD}=2458000$ days and $2459700$ days both consist of a few points from one camera. It is on the upper main sequence and it has a significant IR-excess (Figure \ref{fig:SED_dippers}c) in the 2MASS and \WISE{} photometry compared to an SED model fit to the optical photometry. J094848$-$545959 is similar to J223332$+$565552. \citet{Kuhn_2021} use \textit{Spitzer} photometry to classify this star as a YSO candidate with a disk. 

J162209$-$444247 had a single, deep $\Delta g\simeq0.6$~mag dimming event on 25/04/2023 that lasted for about 3 days. It is located in the red clump. The ASAS-SN light curve also shows evidence for variability on long timescales, similar to the spotted giants in \cite{Petz_2025}. After masking the dimming event, we used a Lomb-Scargle periodogram to identify an 11.4 day period, presumably corresponding to the rotation period of the giant. J162209$-$444247 has also been identified as a coronal line emitter in \textit{eROSITA} \citep{erosita}, with an X-ray luminosity of $(2.32 \pm 0.89)\times10^{31} $ erg s$^{-1}$. There is no \Gaia \space orbital solution, but the large RV amplitude ($RV_{amp}= 94$ km/s) suggests that the system is most likely a long-period, chromospherically active, eclipsing binary. The large $RV_{amp}$ also implies $P=12M_{T\odot}$ days (Eqn.~\ref{eqn:MPtoV}), which means that the observed \Gaia{} RV variability does not correspond to the long-period binary orbit.

J183210$-$173432 had a single, $\Delta g\simeq0.3$~mag dimming event on 02/09/2018 that lasted for about 41 days. It is located on the giant branch. J183210$-$173432 is similar to J223332$+$565552 in shape. 

J183606$-$314826 had a single, $\Delta g\simeq0.4$~mag dimming event on 10/03/2022 that lasted for about 11 days. It is located on the giant branch and is similar to J181752$-$580749. AAVSO and ASAS-SN report J183606$-$314826 as a semi-regular variable with $P=191$ days \citep{aavso, Christy_2023}. The amplitude of the periodic variability at $P=191$~days is smaller than that of the dip. There is a group of bad points from one camera, creating a cloud below the baseline light curve shortly after the event. This is likely due to an intercalibration issue between the fields for that camera. As discussed above, the $RV_{amp}=4.14$~km/s implies an orbit long compared to the $\sim1{,}000$ day span of the \Gaia{} data. This is in agreement with the minimum period estimate of $\sim800$~days.

J190316$-$195739 had a single, $\Delta g\simeq0.4$~mag dimming event on 19/08/2019 that lasted for about 15 days. It is located on the giant branch and is similar to J181752$-$580749. This target has a spectrum from the Radial Velocity Experiment (RAVE) survey \citep{Steinmetz20} that is consistent with a red clump star at an effective temperature $T_{\rm{eff}}=4500$~K and with a surface gravity of $\log g=1.53$. The measured velocity difference between RAVE and \Gaia \space DR3 is $\Delta\rm{RV} = 3.9$~km/s, which is slightly larger than the reported \Gaia \space RV amplitude, $RV_{amp}= 3.4$~km/s. This $RV_{amp}$ implies a long orbit compared to the $1{,}000$ day span of the \Gaia{} data. There are no other indicators of binarity ($\rm{RUWE}=1.09$, $\ipd=0$). 

J205245$-$713514 had a single, $\Delta g\simeq0.4$~mag dimming event on 14/04/2023 that lasted for about 2 days. It is located on the upper main sequence and is similar to J223332$+$565552.

J212132$+$480140 had a single, $\Delta g\simeq0.4$~mag dimming event on 09/09/2021 that lasted for about 58 days. It is located on the giant branch and is similar to J181752$-$580749. \cite{GaiaColab, GaiaDR3} reports the detection of short-period variability ($P=0.0153$ days) for this star, but we do not recover this signal in the ASAS-SN data. The RV amplitude of $RV_{amp} = 1.13$~km/s suggests a long orbit compared to the $1{,}000$ day span of the \Gaia{} data. 

\begin{figure*}
    \centering
    \includegraphics[width=\linewidth]{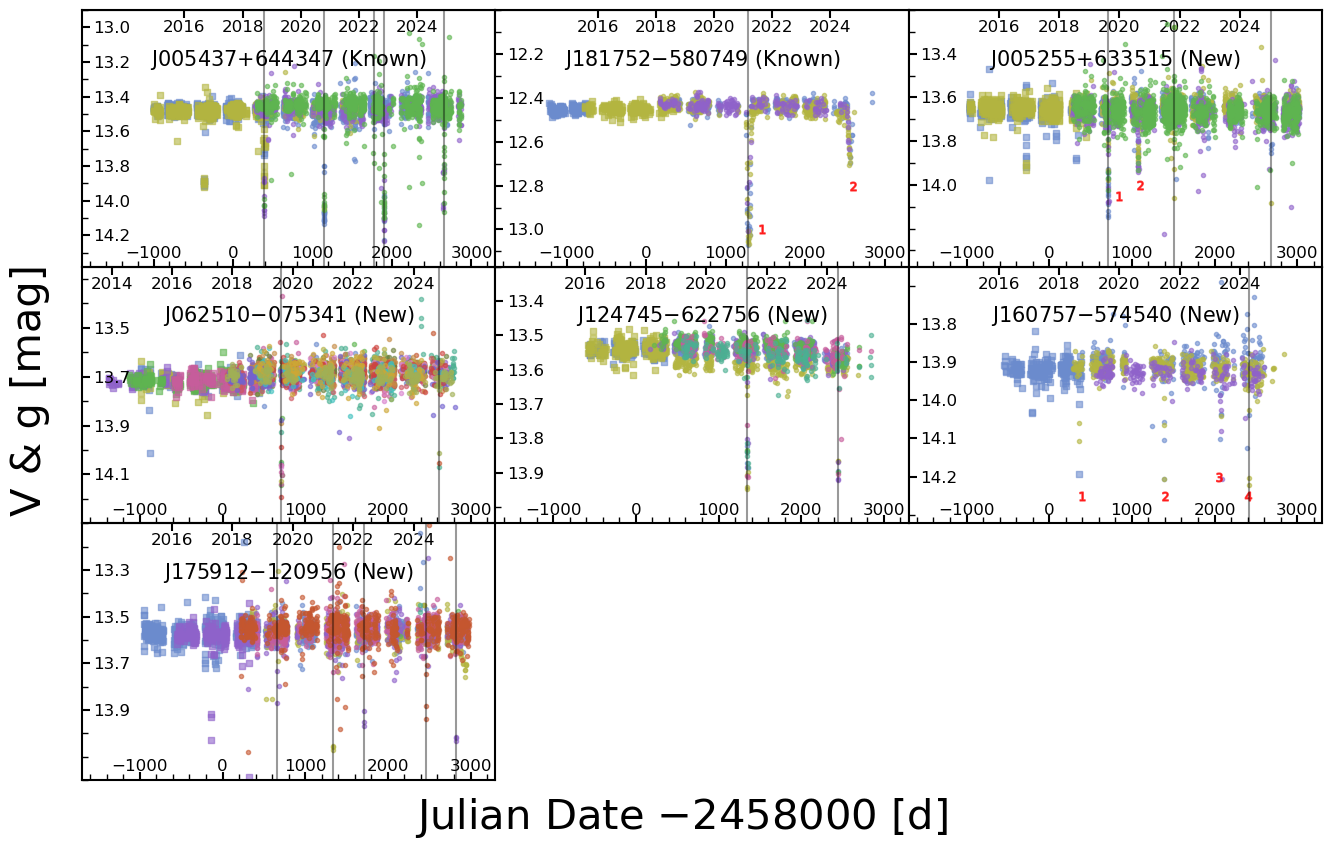}
    \caption{The multi-dip eclipsing binaries. They are organized in the same way as in Figure \ref{fig:SDE_Binaries} and are discussed in Section 3.2. }
    \label{fig:MDE_Binaries}
\end{figure*}

J225702$+$562312 had a single, $\Delta g\simeq0.4$~mag dimming event on 08/06/2019 that lasted for about 6 days. It is located in the red clump and is similar to J175602$+$013135. \Gaia \space reports J225702$+$562312 as a $\beta$ Persei (Algol) eclipsing binary with $P=0.608$ days \citep{GaiaColab, GaiaDR3}. The low $RV_{amp}=13.63$~km/s implies a long orbit compared to the $1{,}000$ day span of \Gaia{}. 

\subsection{Multi-Dip Eclipsing Binaries} 

The multi-dip eclipsing binaries are objects that feature several symmetric dimming events in their ASAS-SN light curves. Similar to the single dip eclipsing binaries, the symmetric shape suggests that a star is the occulting object. 

\begin{figure}[t!]
    \centering
    \includegraphics[width=\linewidth]{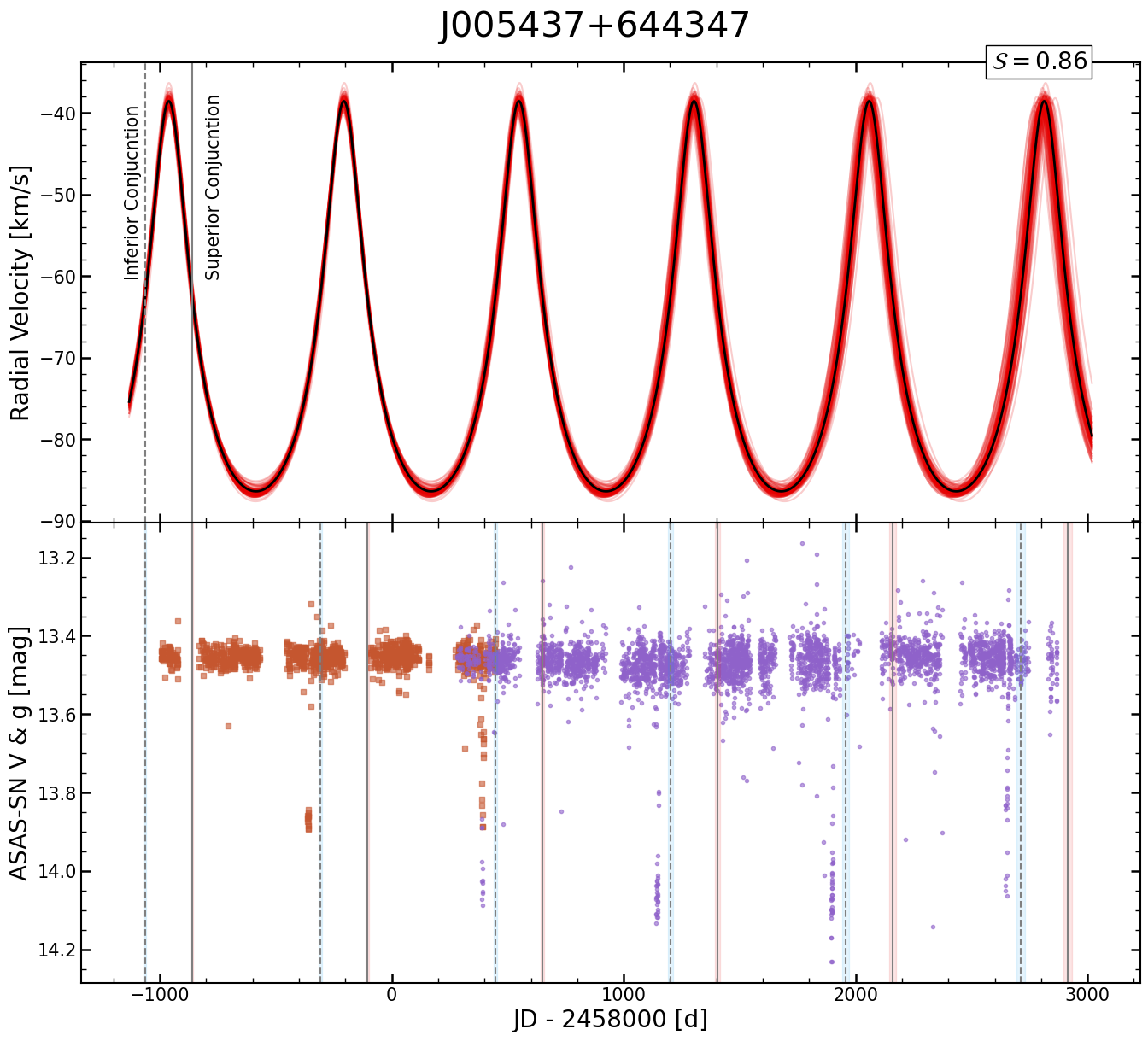}
    \caption{Top: the \Gaia{} RV orbit solution for J005437$+$644347. The black curve corresponds to the median of the reported posteriors, and the red curves show random samples drawn using the covariance matrix of the solution. The vertical lines mark the conjunction times determined using the \Gaia{} orbital solution. Bottom: ASAS-SN light curve of J005437$+$644347. The vertical lines and shaded regions mark the conjunction times extrapolated to the time of the ASAS-SN observations.}
    \label{fig:J0054_SB1}
\end{figure}

J005437$+$644347 (ASASSN-20nq, Figure \ref{fig:MDE_Binaries}, ASAS-SN Transient Page\footnotemark[\ref{fn:transients}]) is on the giant branch and has four distinct dimming events that are identical to each other. The light curve of J005437$+$644347 features an additional vertical line that was triggered by a noisy point in the light curve and is considered a false positive. The dips all reach a depth of $\Delta g\simeq0.6$~mag and last for about 8 days. We found a period of $P=752.3$ days using a Lomb-Scargle periodogram \citep{Lomb_1976,Scargle_1982}. The system has $RV_{amp} =48.8$~km/s and is classified by \Gaia{} as a single lined spectroscopic binary with $P=754.9$ days, $K=23.9\pm0.4$~km/s, consistent with the value of $RV_{amp} \simeq2K$, and $e=0.373$ \citep{GaiaColab, GaiaDR3}. \citet{Bashi22} uses archival RVs and the \Gaia{} RV statistics to predict a score $0\leq S\leq$ for each SB1 orbit. They report a score of $\mathcal{S}=0.86$ for this orbital solution, which is above their recommended minimum of $\mathcal{S}=0.587$ for a good binary solution. For the \Gaia{} orbit, the binary mass function is
\begin{equation}
    f(M) = \frac{PK^3}{2\pi G}(1-e^2)^{3/2} = \frac{M_2^3\sin^3i}{(M_1+M_2)^2}=0.85M_{\odot} 
\end{equation}
and the photometric primary must be at least $M_1\simeq3.4\ M_\odot$ in order for it to be more massive (i.e. more luminous) than the secondary ($M_1>M_2$). J005437$+$644347 sits high on the giant branch, likely implying a high mass. We used the \Gaia{} orbital solution to predict the times of superior, $T_{\rm{sup}}$, and inferior, $T_{\rm{inf}}$, conjunctions. Figure \ref{fig:J0054_SB1} shows the \Gaia{} SB1 orbit and the times of superior and inferior conjunction predicted by the \Gaia{} orbit compared to the ASAS-SN light curve. While the uncertainties on the predicted conjunction times are small ($\lesssim 10$~days), we find that they do not correspond to the observed eclipses in ASAS-SN. If we adjust the \Gaia{} periastron time by $\sim 65$~days to align the inferior conjunction times with the observed ASAS-SN eclipses, the superior conjunctions lie in the seasonal gaps of the ASAS-SN data. The \TESS{} light curve shows a clear eclipse in Sector 58. This corresponds to the third eclipse in the ASAS-SN light curve. This event reaches its lowest point on 15/11/2022. The light curve shows a flat bottom for the eclipse, implying the eclipse is a total eclipse. The eclipse duration of 11 days is consistent with the events in the ASAS-SN data.

J181752$-$580749 \citep[ASASSN-21co,][]{Rowan2021} is on the giant branch and has a single dipping event that began on 10/02/2021 and lasted for about 87 days with a maximum depth of $\Delta g\simeq0.6$~mag. \cite{Way_2021} found an earlier eclipse in ASAS in April 2009 which corresponds to a period of 11.9 years. We later observed a secondary eclipse that began on 11/08/2024 and lasted for about 54 days with a maximum depth of $\Delta g\simeq0.2$~mag. Based on the orbital period, the previous secondary eclipse would have been deepest on 14/10/2012, which is before the first observations of ASAS-SN and after the last observations of ASAS-3 \citep{Pojmanski_2001}. The low value of $RV_{amp}=1.3$~km/s is consistent with \Gaia{} poorly sampling such a long orbit.

\begin{figure*}
    \centering
    \begin{tabular}{ccc}
        \begin{overpic}[width=\panelfigwidth]{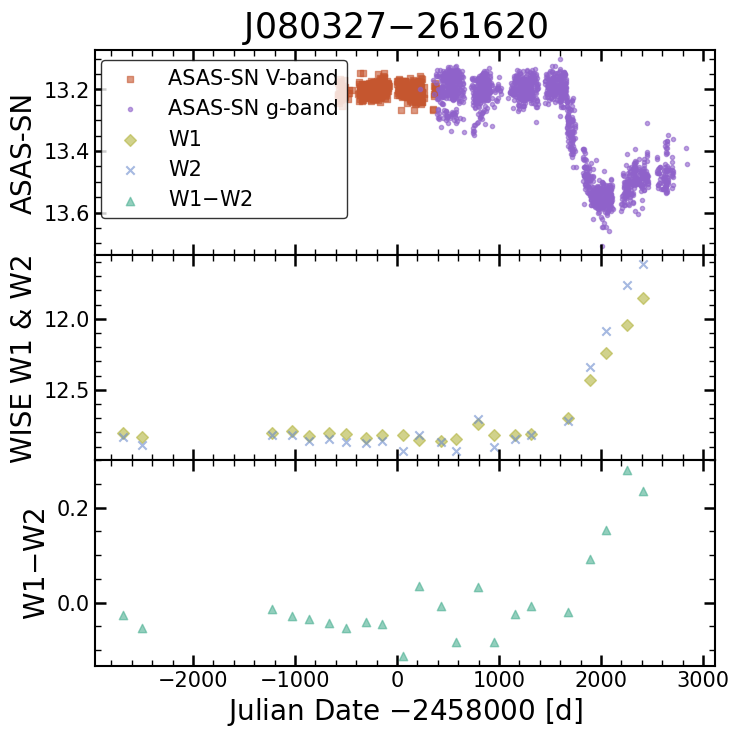}
            \putbox{\xpos}{\ypos}{a}
        \end{overpic} &
        \begin{overpic}[width=\panelfigwidth]{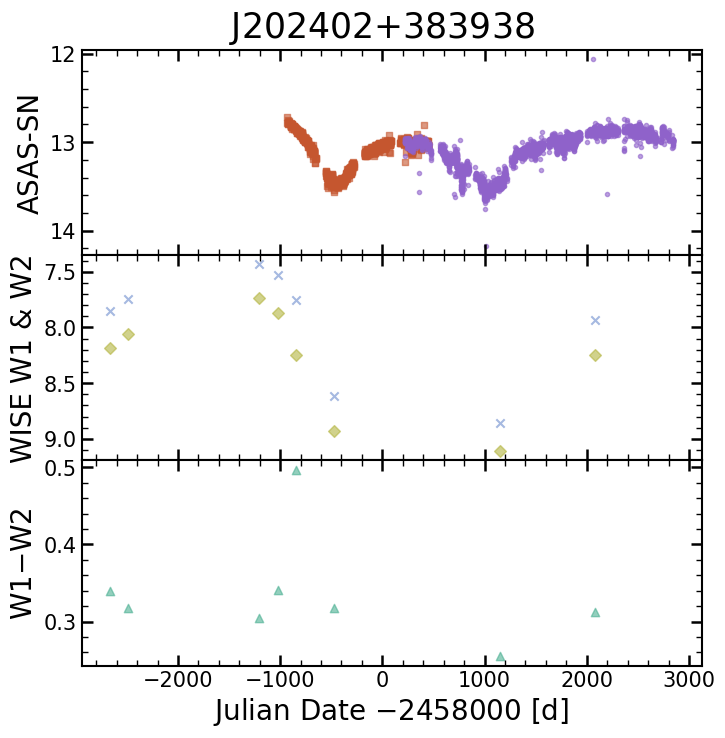}
            \putbox{\xpos}{\ypos}{b}
        \end{overpic} &
        \begin{overpic}[width=\panelfigwidth]{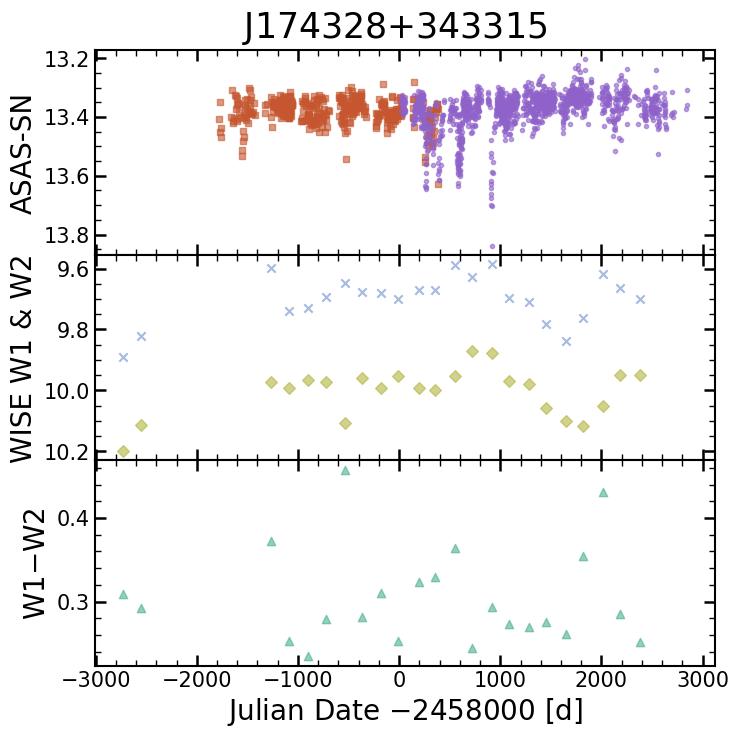}
            \putbox{\xpos}{\ypos}{c}
        \end{overpic} \\
        \multicolumn{3}{c}{
            \begin{tabular}{cc}
                \centering
                \begin{overpic}[width=\panelfigwidth]{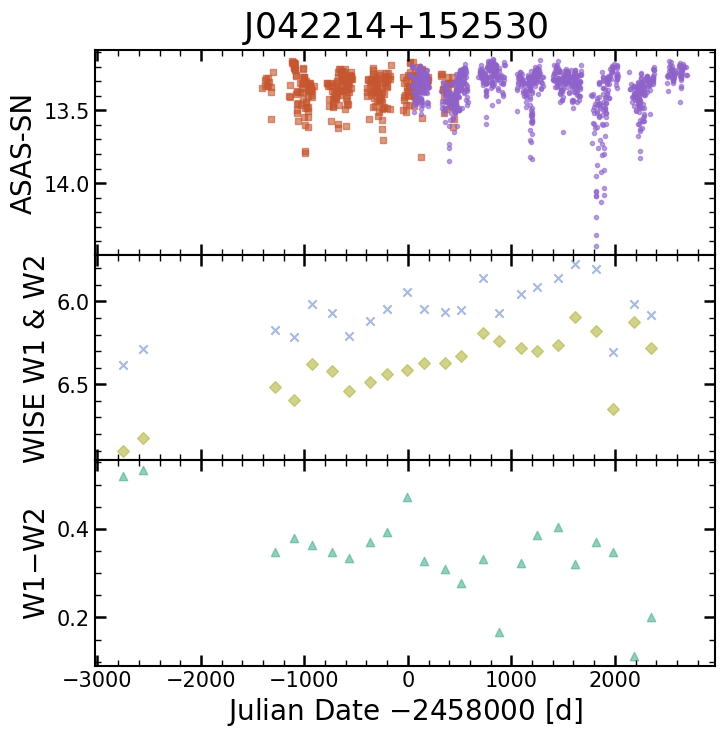}
                    \putbox{\xpos}{\ypos}{d}
                \end{overpic} &
                \begin{overpic}[width=\panelfigwidth]{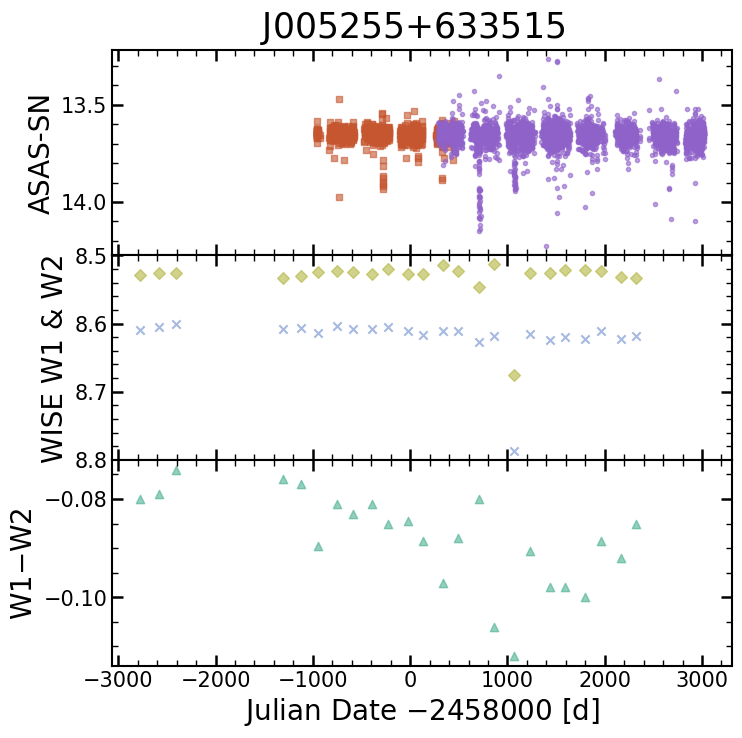}
                    \putbox{\xpos}{\ypos}{e}
                \end{overpic} 
            \end{tabular}
        }\\
    \end{tabular}
    \caption{The \WISE \space light curves of the new dippers with J005255$+$633515. The top panel is the ASAS-SN light curve. The middle panel shows the \WISE \space $W1$ and $W2$ light curves and the bottom panel shows the \WISE \space color. The full ASAS-SN light curve of J005255$+$633515 can be seen in Figure \ref{fig:MDE_Binaries}. The ASAS-SN light curves of the dippers can be seen in Figure \ref{fig:dippers}.}
    \label{fig:WISE_dippers}
\end{figure*}

J005255$+$633515 shows two dimming events (Figure \ref{fig:MDE_Binaries}). The first event began on 11/08/2019 and lasted for 11 days, with a maximum depth of $\Delta g\simeq0.4$~mag. The second event began on 08/08/2020 and lasted for 8 days, with a maximum depth of $\Delta g\simeq0.2$~mag. A dip in the \WISE \space light curve (Figure \ref{fig:WISE_dippers}e, $\Delta W1\simeq0.2$~mag) coincides with the second eclipse. About 986 days before the first event in the $g$-band data, there is the beginning of an eclipse in the $V$-band data on 28/11/2016, with the remainder falling in a seasonal gap. We are inclined to believe these are eclipses, due to the symmetric shape of the events and its detection in multiple cameras. For $P=986$~days, $V_p=21M_{T\odot}^{1/3}$~km/s, so the period seems consistent with the \Gaia{} radial velocity amplitude of $RV_{amp} = 26.55$~km/s. J005255$+$633515 is on the giant branch and is similar to J181752$-$580749.

J062510$-$075341 had two deep dimming events and the first began on 14/08/2019 and lasted for about 4 days, with a depth of $\Delta g\simeq0.7$~mag. The second event began on 30/10/2024 and lasted for 3 days, with a depth of $\Delta g\simeq0.5$~mag. The minima of the two events are separated by about 1905 days. The target is a red giant and the event is similar to the event seen in J005437$+$644347 above. \Gaia{} reports a single-lined spectroscopic binary (SB1) orbit solution with an orbital period of $P=1134 \pm 285$ days, a velocity semi-amplitude of $K=23.4\pm4.7$~km/s, and an eccentricity of $e=0.32\pm0.20$ \citep{GaiaColab, GaiaDR3}. This target has a binary orbit score of $\mathcal{S}=0.66$, which is above the minimum recommended for a good orbital solution by \cite{Bashi22}. Note that the period of $P=278 M_{T\odot}$~days, estimated from Eqn~\ref{eqn:MPtoV} and $RV_{amp}=23.67$~km/s is consistent with the SB1 period, as well as the period seen in the ASAS-SN light curve, if the mass is fairly high. J062510$-$075341 sits on the giant branch above the red clump, but it is hard to determine the masses of giants. We use the \Gaia{} model to predict that the expected Julian Dates of the superior and inferior conjunctions are $T_{\rm{sup}}=2457070^{+60}_{-70}$ and $T_{\rm{inf}} = 2456730^{+100}_{-200}$. However, because the \Gaia{} RVs were taken $>2$ years prior to the observed ASAS-SN event, and the \Gaia{} period uncertainty is $\sigma_P=285$~days, we cannot make a meaningful comparison between the conjunction times predicted by the \Gaia{} orbit and the observed eclipse. The next \Gaia{} data release is expected to include epoch radial velocity measurements, and a simultaneous fit to the \Gaia{} RVs and the ASAS-SN light curve will allow the orbital elements to be better constrained.

J124745$-$622756 shows two deep, dimming events. The first event begins on 10/05/2021 and lasts for about 7 days with a maximum depth of $\Delta g\simeq0.4$~mag. The second event begins on 14/05/2024 and also lasts for about 7 days with the same maximum depth. The two dips are separated by about 1096 days. This target is similar to the \Gaia \space single lined spectroscopic binary J005437$+$644347 (Figure \ref{fig:MDE_Binaries}). J124745$-$622756 is in the Hertzsprung Gap, but the star is in the Galactic plane ($b=0.403$ deg) and the 3-dimensional dust maps predict a high extinction ($A_G = 3.61$~mag), so the CMD position is uncertain. It has a radial velocity amplitude of $RV_{amp} = 54.89$ km/s in \Gaia \space \citep{GaiaColab, GaiaDR3}, while for a 1096 day period, $V_p=21M_{T\odot}^{1/3}$~km/s. The high velocity amplitude $RV_{amp}$ implies a high mass or eccentricity. The first eclipse egress is seen in the \TESS{} sector 38 light curve at the expected time.

J160757$-$574540 is a red clump star with three distinct deep, dimming events in the $g$-band, with one proceeding $V$-band. The first dimming event began on 26/08/2018, with a depth of $\Delta V\simeq0.3$~mag, and lasting for about 3 days. The second dimming event began on 20/06/2021, reached a depth of $\Delta g\simeq0.3$~mag, and lasted for about 4 days. The third dimming event began on 04/05/2023, had a depth of $\Delta g\simeq0.3$~mag, and lasted for about 8 days. The fourth dimming event began on 17/04/2024, reached a maximum depth of $\Delta g\simeq0.3$~mag and lasted for about 2 days. The minima of the first and second event are separated by about 1031 days. The minima of the second and third events are separated by about 706 days, while the minima of the primary eclipses (second and fourth events) are separated by about 1030 days. In total there are four dimming events, but the primary and secondary eclipses are not separated by 515 days, indicating that the binary has a non-zero eccentricity. This target has a radial velocity amplitude of $RV_{amp} = 36.67$ km/s in \Gaia \space \citep{GaiaColab, GaiaDR3} which suggests binarity. A 1030 day period implies $V_p=21M_{T\odot}^{1/3}$~km/s which requires the source to have either a high mass or eccentricity to have the significantly larger $RV_{amp}$. J160757$-$574540 sits on the giant branch above the red clump, but is difficult to infer the masses for giants. The second and third eclipses were partially observed by \TESS{} in sectors 39 and 65, respectively.

J175912$-$120956 features four deep, dimming events. The first identified event is a false positive due to noisy data. The first real $g$-band event takes place on 06/05/2021 with a maximum depth of $\Delta g\simeq0.5$~mag and lasts for less than a day. The second event takes place on 12/05/2022 with a maximum depth of $\Delta g\simeq0.4$~mag and, also lasts for less than a day. The minima of the dips are separated by about 371 days. The third $g$-band takes place on 28/05/2024 with a maximum depth of $\Delta g\simeq0.4$~mag, also lasts for less than a day, and has a separation of 747 days from the previous event. The fourth $g$-band takes place on 30/05/2025 with a maximum depth of $\Delta g\simeq0.5$~mag, also lasts for less than a day, and has a separation of 367 days from the previous $g$-band event. There was also a dimming event in the $V$-band that began on 12/04/2017 with a maximum depth of $\Delta V\simeq0.4$~mag and also lasting less than a day. There appears to be a combined $V$-band/$g$-band event after this $V$-band event, however, it is not real and consists of several points at the mean magnitude of the light curve in between the apparent points of the event. The nonphysical shape of this event leads us to believe it is a false positive. The minima of the $V$-band event and the first $g$-band event are separated by about 1485 days. This is 4 times 371, so this is a likely eclipsing binary that has a period of $P=371$ days, consistent with the separations of the $g$-band dips. Even though the duration of the eclipses is short, we detect the eclipses in multiple cameras. The first $g$-band event has two points from different cameras. The second $g$-band event has six points from two different cameras. The \TESS{} light curve shows the dip in sector 92 corresponding to the fourth $g$-band dip in the ASAS-SN light curve. Its short duration of $<1$ days is consistent with the duration of ASAS-SN results. This target has a radial velocity amplitude of $RV_{amp} = 204.92$ km/s in \Gaia{} \citep{GaiaColab, GaiaDR3}. With a period of $P=371$ days, $V_p=30M_{T\odot}^{1/3}$~km/s. The large $RV_{amp}$ cannot be due to the long-period orbit, but it is possible that this system is instead a hierarchical triple, with the $RV_{amp}$ corresponding to an inner binary and the eclipses being due to the more distant companion. However, J175912$-$120956 is on the upper main sequence, and since Gaia RVs can be poorly measured for hot stars, it is also possible that the RV amplitude estimate is biased.

\subsection{Dippers}

\begin{figure*}
    \centering
    \includegraphics[width=\linewidth]{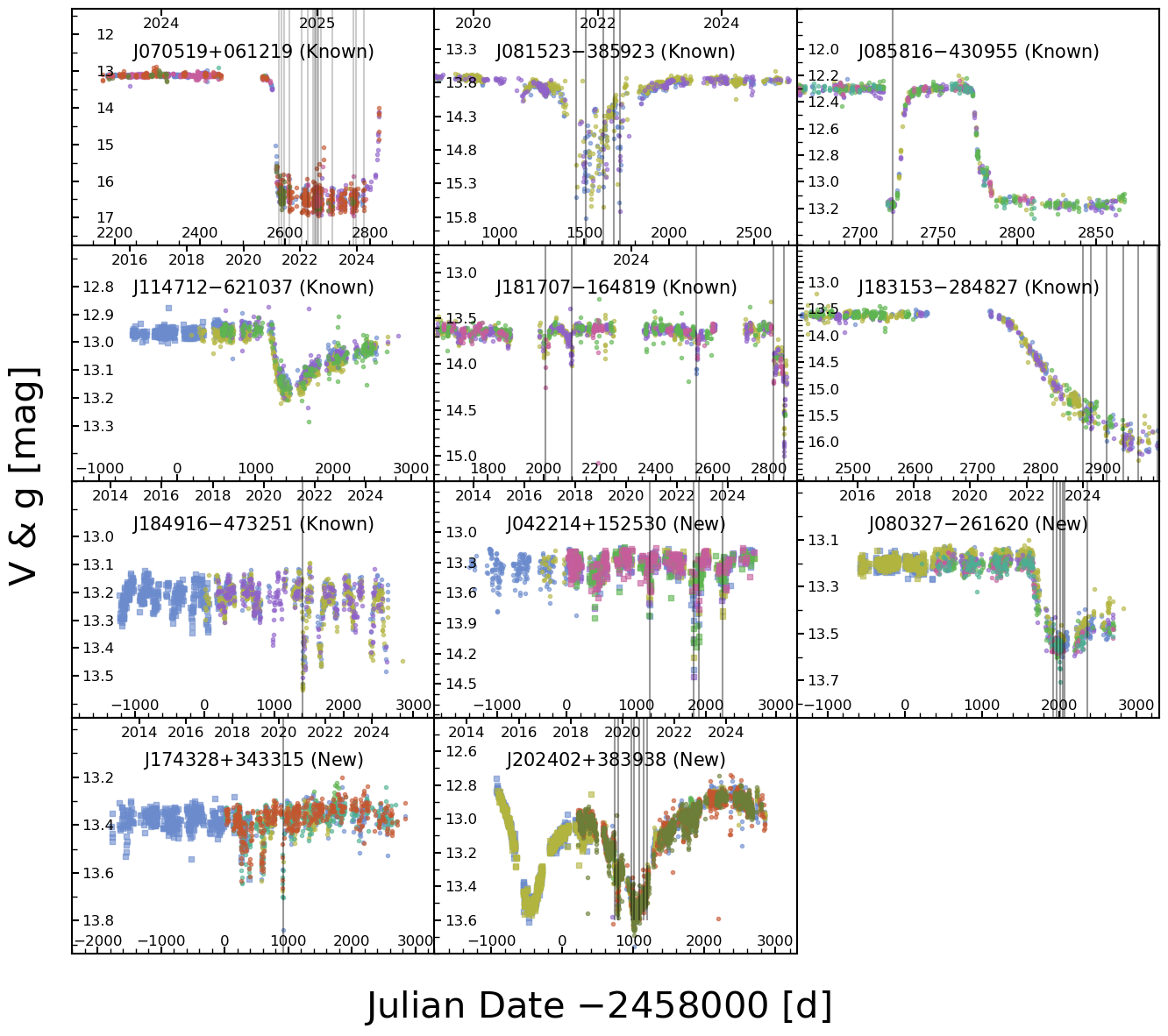}
    \caption{The dippers discussed in Section 3.3, organized in the same way as Figures \ref{fig:SDE_Binaries} and \ref{fig:MDE_Binaries}. Table \ref{table:dip_param} reports the number of detected peaks and the properties of the events.}
    \label{fig:dippers}
\end{figure*}

Dipper objects are a wide category of extrinsic stellar variables, some of which can exhibit long and deep events. They are usually classified into two classes ($\epsilon$ Aurigae and KH 15D) although there are other configurations capable of producing dips with many different types of features such as occulations produced by debris fragments \citep[see, e.g., Boyajian's Star,][]{Boyajian_2016, Simon_2018} or ring like structures \citep[see, e.g.,][]{Mamajek_2012}. Many of the dippers found serendipitously and through our search are caused by dust, as can be seen from the IR-excesses of some of the objects. 

J070519$+$061219 \citep[ASASSN-24fw, Figure \ref{fig:dippers},][]{JoHantgen_2024_2} is an upper main sequence star with an 8 month long, deep ($\Delta g =4.1$~mag), dimming event that began on 16/09/2024 and ended in early June 2025. This event was the subject of an extensive follow-up campaign and is characterized in detail in \cite{Fores-Toribio_2025} and \cite{Zakamska_2025}. It was caused by dust obscuring the object, and it has a period of 43.8 years \citep{Nair_2024, Fores-Toribio_2025, Zakamska_2025}.

J081523$-$385923 \citep[ASASSN-21qj,][]{Rizzo_Smith_2021_qj, Kenworthy2023, Marshall_2023} is a lower main sequence star with a deep, long-lasting dimming event that began on 30/03/2021 and lasted for about 608 days. The dip reached a maximum depth of $\Delta g\simeq2.0$~mag. \cite{Kenworthy2023} and \cite{Marshall_2023} both explain the event as the obscuration due to a dust cloud created in a recent collision based on the mid-IR brightening observed in the period before the dimming event begins. 

J085816$-$430955 \citep[ASASSN-25bv,][]{JoHantgen_2025} is an upper main sequence star with two deep dimming events that are separated by about 29 days. The first dip began on 14/02/2025, lasted for roughly 27 days and reached a maximum depth of $\Delta g\simeq0.9$~mag. The shape of this event was not symmetric, and the egress was also captured in \TESS{} Sector 89. After the minimum, there is a slow rise in flux until 06/04/2025, when it began to fade again and plateaued at a depth of $\Delta g\simeq0.6$~mag from 10/04/2025 to 16/04/2025 before fading by an additional 0.3 mag. The event is still ongoing. This star has a RUWE = 5.44, and the \Gaia \space $RV_{amp} = 24.85$~km/s, suggesting it is a part of a binary system \citep{GaiaColab, GaiaDR3}. The $RV_{amp}$, suggests a possible period of $P=630M_{T\odot}$~days. The shape of the light curve is qualitatively similar to massive stars occulted by disks, such as TT Nor \citep{Rowan_2023_2}, but the system would have a much longer period. We looked at the DASCH light curve \citep{Laycock_2010, Grindlay_2012, Tang_2013} to try to constrain the period, but there is too much scatter in the data to search for earlier events.

J114712$-$621037 \citep[ASASSN-21js,][]{Pramono_2024} is an upper main sequence star with an ongoing, large, deep dimming event that began on 20/11/2020. The maximum depth so far is $\Delta g\simeq0.3$~mag. \cite{Pramono_2024} explained the event as being due to a large dust ring occulting the star and predicted the event will end in January 2027.

J181707$-$164819 (ASASSN-25cy, ASAS-SN Transient Page\footnotemark[\ref{fn:transients}]) is an upper main sequence star with several, irregularly spaced, deep dimming events. The most recent event began on 11/05/2025 and it has faded to a maximum depth of $\Delta g\simeq1.4$~mag. This object appears in Be star catalogs \citep{Vioque_2020} and is a YSO candidate \citep{Marton_2016}. 

J183153$-$284827 \citep[ASASSN-25cd,][]{JoHantgen_2025_2} is a red giant star with a deep dimming event that began on 30/03/2025 and is still fading. It reached a depth of $\Delta g\simeq2.5$~mag as of 15 January 2026. Broadband photometry of this target before the dimming is consistent with a $L=40L_{\odot}$ K-giant \citep{McCollum_2025}. As discussed above, the $RV_{amp}=2.85$~km/s implies an orbit long compared to the $1{,}000$ day span of the \Gaia{} data.

J184916$-$473251 \citep[ASASSN-21nn,][]{Rizzo_Smith_2021_nn} is on the giant branch and has a complex light curve with a single deep dimming event that reached a depth of $\Delta g\simeq0.3$~mag. The event began on 08/07/2021 and lasted for about 75 days. The star also shows periodic variability with a period of about $P=232$ days. \cite{Buckley_2021} reports a double peaked H$\alpha$ "shell line" where a rotationally broadened photospheric line is observed in the presence of an edge on disk around the star \citep[see, e.g.,][]{Rivinius_2006}. The $RV_{amp}=6.29$~km/s implies an orbit long compared to the $1{,}000$ day span of the \Gaia{} data.

J042214$+$152530 shows several dimming events that are irregularly spaced in time. The deepest event begins on 20/07/2022 and lasts for about 135 days. It reaches a maximum depth of $\Delta g\simeq1$~mag. This target is similar to J005437$+$644347 (ASASSN-20nq). J042214$+$152530 was classified as a YSO by ASAS-SN \citep{Jayasinghe18}, a "dubious" variable by ATLAS with a period of $P=433.6$ days \citep{Heinze18}, and a rotational variable by ZTF \citep{Qiao24}. There are two archival LAMOST spectra, taken in on 2016-02-05 and 2017-01-28, with similar RVs ($17\pm8$~km/s and $15\pm7$~km/s, respectively). The extinction corrected color, $B_{\rm{P}}-R_{\rm{P}} = 1.17$~mag and absolute magnitude, $M_G=3.5$, places it just below the giant branch (Figure \ref{fig:cmd}). The SED (Figure \ref{fig:SED_dippers}d) has an IR excess in the 2MASS and \WISE{} filters and the target is brighter in the $W4$ band than in the $W3$ band. The target has also gotten brighter in the $W1$ and $W2$ filters over the past $\sim13$ years with $\Delta W1 \simeq 0.8$~mag, followed by a dip ($\Delta W1 \simeq 0.7$~mag) coinciding with the optical dip (Figure \ref{fig:WISE_dippers}d). \textit{Chandra} \citep{Evans_2024} reports an X-ray detection with a luminosity of $(3.96 \pm 2.36)\times10^{28}$ erg $s^{-1}$ using the \Gaia{} distance, which is typical of YSOs \citep{Forbrich_2011}. The $RV_{amp}=54.18$~km/s suggests a period of $P=61M_{T\odot}$~days. 

J080327$-$261620 \citep[ASASSN-24fa,][]{JoHantgen_2024} shows a long-lasting, deep dimming event that began on 31/01/2021 and reached a maximum depth of $\Delta g\simeq0.5$~mag on 07/03/2023. Since then the flux has been slowly recovering. We do not see any earlier events. J080327$-$261620 is on the upper main sequence and is similar to J114712$-$621037. J080327$-$261620 is reported by \Gaia \space as a short period variable \citep[$P=0.461$ days,][]{GaiaColab, GaiaDR3}, while ATLAS assigns it a low confidence period of $P=0.921$ days \citep{Heinze18}, which is roughly twice as long as the \Gaia{} period. AAVSO identifies it as $\gamma$ Cassiopeiae or a $\lambda$ Eri type variable with the \Gaia \space period. There is also a mid-IR excess (Figure \ref{fig:WISE_dippers}a) that begins when the optical light fades and is steadily increasing ($\Delta W2\simeq1$~mag). The \WISE{} color also gets redder as the event progresses, suggesting that this event is caused by newly formed dust. 

\begin{figure}
    \centering
    \includegraphics[width=\linewidth]{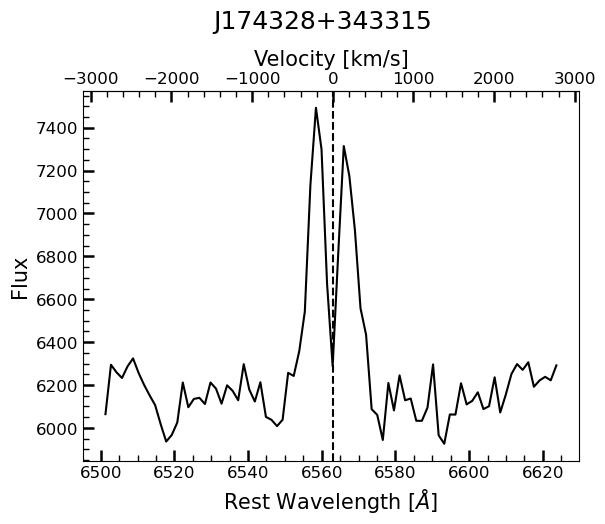}
    \caption{The LAMOST H$\alpha$ line showing its double-peaked H$\alpha$ emission.}
    \label{fig:Ha}
\end{figure}

J174328$+$343315 shows several deep, dimming events around 22/09/2017 and they continue in both the $g$-band data and $V$-band data for about 1000 days. These dimming events share some similarities with J184916$-$473251 (ASASSN-21nn, Figure \ref{fig:dippers}). This star was classified as a YSO in ASAS-SN \citep{Jayasinghe18}, a ``dubious'' variable in ATLAS \citep{Heinze18}, a suspected variable star in ZTF \citep{ztf}, and as an RS CVn variable in \Gaia{} \citep{GaiaColab, GaiaDR3}. Its CMD position in the Hertzsprung Gap is atypical for RS CVn stars. Since Hertzsprung gap stars are rare, its CMD location may imply that it is an unresolved binary with a red primary and a blue secondary. Its parallax uncertainty is small ($\varpi = 0.5140 \pm 0.0177$ mas) and the extinction is only $A_G = 0.19$~mag, suggesting the CMD position is correct. It was observed by LAMOST \citep[LAMOST-LRS]{Cui12}, and the spectrum shows a double-peaked H$\alpha$ emission line (Figure \ref{fig:Ha}). This, combined with the \WISE{}  infrared excess and variability seen in Figures \ref{fig:SED_dippers}a and \ref{fig:WISE_dippers}c, suggest that there is an accretion disk obscuring the star and producing irregular variability \citep{Carroll_1991}. In \WISE{}, both the W1 and W2 light curves fade by $\simeq 0.2$~mag following the most recent event. The source has an $RV_{amp}=84.84$~km/s, implying a period of $P=16M_{T\odot}$~days.

\begin{figure*}
    \centering
    \includegraphics[width=0.8\textwidth]{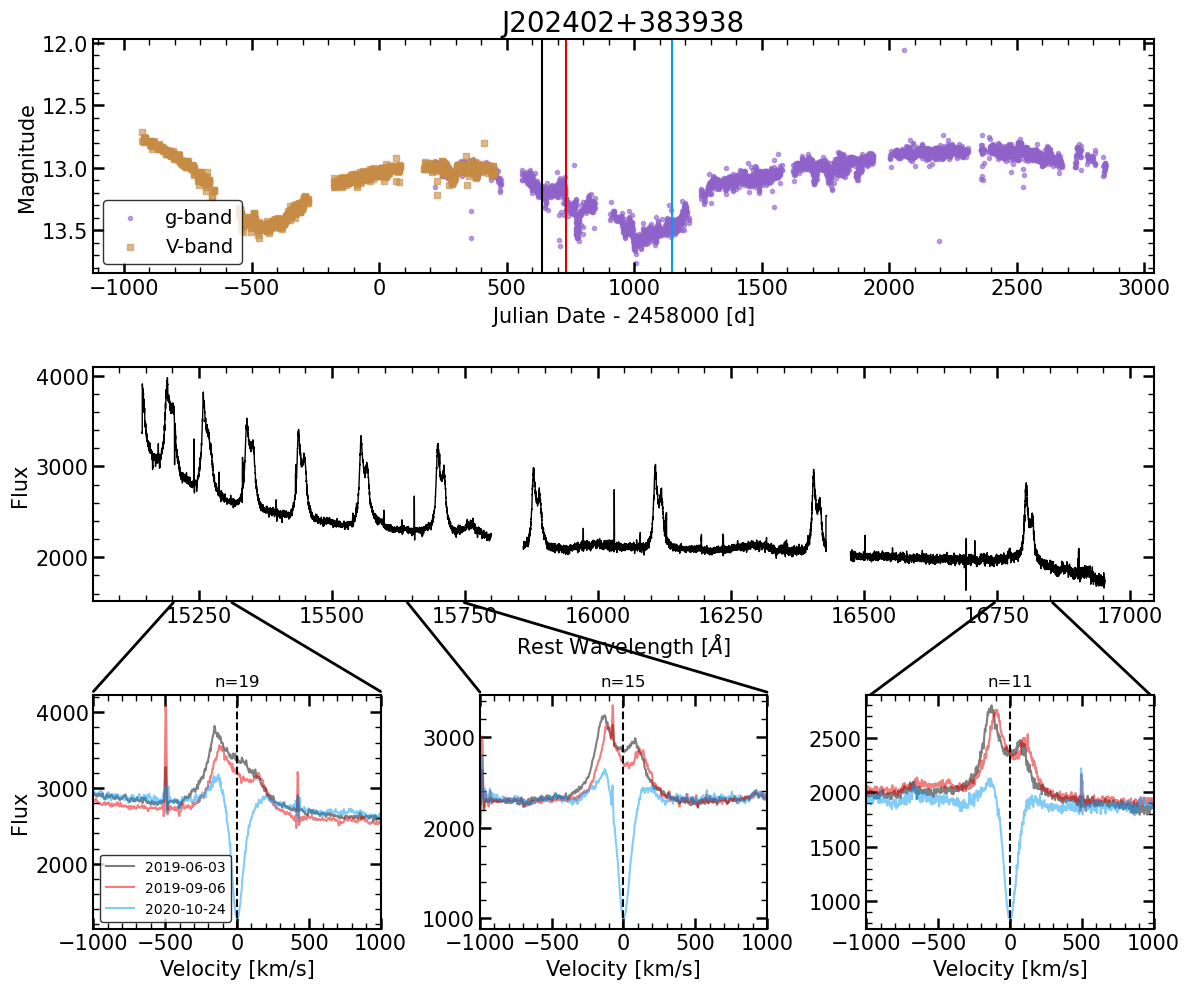}
    \caption{Rest frame spectra of J202402+383938. The top panel shows the full ASAS-SN light curve of J202402+383938, with vertical lines at the epochs of the near-IR spectra. The middle panel shows the full spectrum at the first (black line) epoch on 03/06/2019. It is dominated by Hydrogen Brackett emission lines. The lower panels show the n = 19 (left), 15 (middle), and 11 (right) Brackett lines at the three spectral epochs with the same color coding as in the upper panel. The lines change from broad double peaked emission lines in the first two epochs to an absorption dominated profile in the last epoch.} 
    \label{fig:J2038}
\end{figure*}

J202402$+$383938 has a long-lasting, deep, dimming event. The light curve began fading on 14/04/2018 and reached a maximum depth ($\Delta g\simeq0.8$~mag) on 14/07/2020. The flux has been slowly recovering since then. The event is asymmetric and includes a number of sharp features which were separately identified by our search. The first sharp feature appears to be a smaller dipping event beginning on 16/10/2019 and reached a maximum depth ($\Delta g\simeq0.4$~mag) on 22/10/2019. There was an earlier deep, dipping event in the ASAS-SN $V$-band data (top panel of Figure \ref{fig:J2038}). The minima of the two events are separated by about 1,470 days. J202402$+$383938 is classified in ATLAS as a long-period variable with $P=986$ days \citep{Heinze18}. J202402$+$383938 is in the Hertzsprung Gap. The spectral energy distribution shows an infrared excess in W1 -- W4 (Figure \ref{fig:SED_dippers}b) and the \WISE{} fluxes also drop by $\Delta W1 \simeq1.4$ mag ($\Delta W2 \simeq1.3$ mag) during the event (Figure \ref{fig:WISE_dippers}b). This suggests the event is an eclipse by a circumstellar disk composed of large particles. J202402$+$383938 was classified as a Brackett-emission line star in APOGEE \citep{Campbell23}. One of the three APOGEE spectra is shown in the middle panel of Figure \ref{fig:J2038}. It has many broad, double-peaked Brackett emission lines. The first two spectra were taken as the star dimmed and the third spectrum was taken as it brightened (top panel of Figure \ref{fig:J2038}). The bottom panel zooms in on the Brackett transitions of the spectra corresponding to $n = 11$, $15$, and $19$. The first two spectra show broad, double-peaked emission lines consistent with the infrared spectra of Be stars \citep{Huang_1972}. In the third spectra, the emission lines are replaced by absorption lines, similar to those seen in the spectra of shell stars \citep{Rivinius_2006}. The emission lines of the first two spectra match that of a disk seen edge-on \citep{Waters_1992}. The emission lines likely come from the inner disk, the magnetosphere of the star and the disk wind region of the star \citep{Tambovtseva_2016, Tambovtseva_2017}. 

\section{Discussion} \label{sec:Conclusions from the targets}

The "occasionally" dimming objects found by ASAS-SN are long-period eclipsing binaries (e.g. ASASSN-23ht and ASASSN-20nq), or occultations by dusty material (e.g. ASASSN-24fw). Long-period eclipsing binaries, specifically those on the red giant branch, can be used to make precise mass and radius measurements \citep[see, ][]{Rowan_2025}. Dust occulters, like ASASSN-21qj, are probes of dusty disks and collisional dust formation events in stellar systems \citep[see, ][]{Kenworthy2023, Marshall_2023}.

Here, we carried out a systematic search for drops in flux $\Delta g > 0.3$~mag that were detected in multiple ASAS-SN cameras for stars with $13 < g < 14$ and light curve dispersions $< 0.15$~mag. The main sources of false positives were stars affected by nearby saturated stars and higher amplitude variable stars (Figure \ref{fig:False_Positives}). Of the previously known systems in ASAS-SN only J114712$-$621037 is missed because it is below our amplitude threshold ($\Delta g<0.3$~mag) and spans a long time period \citep[$1{,}000$~days,][]{Pramono_2024}. 

\begin{figure*}
    \centering
    \includegraphics[width=\textwidth]{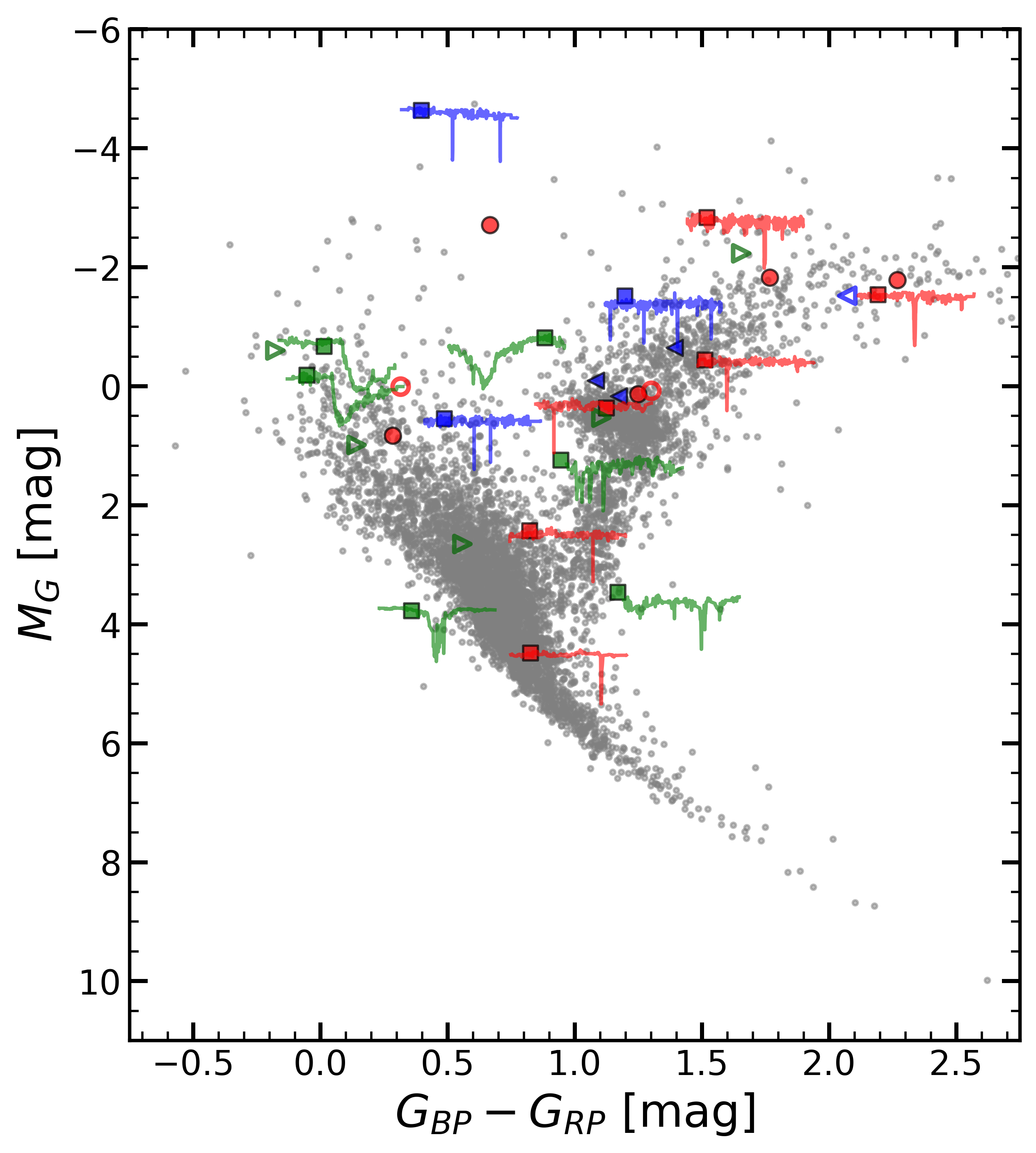}
    \caption{CMDs of each the new and known sources. The light curves are overlaid to compare the various different types of events. Some light curves are omitted to properly display the shown light curves without overlaying them upon each other. The CMD locations of these light curves are indicated by the squares touching the light curve. The rest of the sources are presented similarly to Figure \ref{fig:cmd}, the SDEs are red, the MDEs are blue, and the dippers are green.}
    \label{fig:LC_CMDs}
\end{figure*}

After visually inspecting the candidates, we were left with 15 new (5 known) long period EBs (Figures \ref{fig:SDE_Binaries} and \ref{fig:MDE_Binaries}) and 4 new (7 known) dippers (Figure \ref{fig:dippers}). There is some ambiguity in the objects we classified as single dimming-event long period EBs as compared to dippers. For the 7 EBs where we could estimate periods, we found periods ranging from 371 days to 11.9 years. The dippers are a combination of YSOs, a Be star, and disk occultations. The most striking example of a dipper is J070519$+$061219 (ASASSN-24fw), which had an 8 month long, 4~mag deep occultation due to dusty structures and is analyzed in detail in \cite{Fores-Toribio_2025} and \cite{Zakamska_2025}. The light curves of several of these targets are compared on a \Gaia{} CMD in Figure \ref{fig:LC_CMDs}, which shows the dramatic variation in light curve shape and CMD position of these objects.

Here, we only searched for long-period binaries and dipper stars within the magnitude range of $13<g<14$~mag. Additional searches across the entire ASAS-SN magnitude range ($11<g<16$~mag) would certainly detect many more sources. This search is also complementary to the search for slowly varying ASAS-SN stars in \cite{Petz_2025} and \cite{Petz_2025_2}. There, seasonal medians were used to eliminate shorter time scale variability and then linear and quadratic fits to the medians were used to identify sources slowly varying in brightness. Of the sources here, J080327-261620 (ASASSN-24fa) was also flagged in \cite{Petz_2025}.

There are multiple improvements that can be made to our approach. First, we can identify stars affected by nearby saturated stars in a more automated manner so that fewer sources require visual inspection. Second, we could flag out known variables with well defined types (e.g. EBs). Here we did not do so in order to get a better sense of what our search would find. In any new search, we would simply remove these automatically. Third, we can improve the method to filter out typical stellar variability. Our simple light curve dispersion limit ($\sigma>0.15$) can eliminate longer, deeper events because they are large enough to affect the dispersion while also not eliminating high amplitude variable stars that trigger our simple approach to flagging candidates. One improvement would be to look for both upward and downward dips about the median of the light curve. Variable stars would show both brightening and dimming features, while our desired targets would primarily feature dimming events. We can also better automate the identification of bad data points (see Figure \ref{fig:False_Positives}), and remove them before beginning the search. These approaches should then allow searches with a lower threshold on $\Delta g$ across a wider range of apparent magnitudes.

\begin{table*}
    \centering
    \caption{The objects sorted by discovery and as single-dip eclipsing binaries, multi-dip eclipsing binaries, and dippers. Each target has the source id, the right ascension, the declination, the mean apparent $g$-magnitude, the mean extinction-corrected \Gaia{} absolute $G$-magnitude and color, the radial velocity amplitude ($RV_{amp}$), the RUWE value, and the distances from \cite{Bailer-Jones_2021}.}
    \sisetup{table-auto-round,
     group-digits=false, separate-uncertainty = true, retain-unity-mantissa = false, table-number-alignment = center,
     table-figures-uncertainty = 1, table-parse-only = false, multi-part-units = brackets, parse-numbers = true}
    \setlength{\tabcolsep}{8pt}
    \renewcommand{\arraystretch}{1.7}
    \begin{tabular}{l S[table-format=3.5] S[table-format=3.5] l S[table-format=2.2] S[table-format=2.2] S[table-format=2.2] S[table-format=3.2] S[table-format=3.2] r@{}l}
        \toprule
        {Source} & 
        {RA} & 
        {DEC} & 
        \multicolumn{1}{c}{\shortstack{{Search} \\ {Method}}} & 
        {Mean $g$} & 
        \multicolumn{1}{c}{\shortstack{{Mean Absolute} \\ {$G$}}} & 
        {$G_{\rm{BP}} - G_{\rm{RP}}$} & 
        {$RV_{amp}$} &
        {RUWE} &
        {Distance} \\
        {} & {[deg]} & {[deg]} & {} & {[mag]} & {[mag]} & {[mag]} & {[km/s]} & {} & {[pc]} \\
        \midrule
        \multicolumn{10}{l}{\textbf{Single Eclipse Binaries}} \\
        J073234$-$200049 & 113.142375 & -20.01358333 & Known & 14.48 & 0.01 & 0.32 &  & 1.065 & $3646$ & $^{+254}_{-284}$ \\
        J175602$+$013135 & 269.0079973 & 1.52628966 & Known & 14.53 & 0.08 & 1.30 &  & 1.026 & $3361$ & $^{+176}_{-188}$ \\
        J223332$+$565552 & 338.384125 & 56.93108333 & Known & 12.74 & 4.49 & 0.83 & 22.29 & 0.879 & $354$ & $^{+1}_{-1}$\\
        J073924$-$272916 & 114.8503576 & -27.48776143 & Pipeline & 13.003 & -2.838 & 1.521 & 23.91 & 1.178 & $5192$ & $^{+398}_{-332}$ \\
        J074007$-$161608 & 115.0276126 & -16.26896725 & Pipeline & 13.104 & -2.712 & 0.666 &  & 2.284 & $8202$ & $^{+1484}_{-1391}$ \\
        J094848$-$545959 & 147.1987289 & -54.99963893 & Pipeline & 13.065 & 0.826 & 0.284 &  & 1.048 & $1711$ & $^{+43}_{-38}$ \\
        J162209$-$444247 & 245.5357562 & -44.71308061 & Pipeline & 13.957 & 0.128 & 1.250 & 93.90 & 0.965 & $1772$ & $^{+51}_{-63}$ \\
        J183210$-$173432 & 278.0424857 & -17.57545433 & Pipeline & 12.722 & -1.783 & 2.268 &  & 0.953 & $1942$ & $^{+123}_{-85}$ \\
        J183606$-$314826 & 279.0252466 & -31.80714478 & Pipeline & 13.866 & -1.825 & 1.766 & 4.14 & 0.896 & $6261$ & $^{+565}_{-619}$ \\
        J190316$-$195739 & 285.8184954 & -19.9609327 & Pipeline & 13.608 & -0.4437 & 1.512 & 3.4 & 1.088 & $2931$ & $^{+84}_{-111}$ \\ 
        J205245$-$713514 & 313.1890422 & -71.58708665 & Pipeline & 13.410 & 2.431 & 0.823 &  & 1.346 & $1224$ & $^{+33}_{-30}$ \\
        J212132$+$480140 & 320.3840472 & 48.02785275 & Pipeline & 13.147 & -1.5386 & 2.193 & 1.13 & 0.854 & $1528$ & $^{+65}_{-59}$ \\
        J225702$+$562312 & 344.2585271 & 56.38663345 & Pipeline & 13.827 & 0.3645 & 1.124 & 13.63 & 1.139 & $2249$ & $^{+67}_{-80}$ \\
        \addlinespace
        \multicolumn{10}{l}{\textbf{Multiple Eclipse Binaries}} \\
        J005437$+$644347 & 13.652625 & 64.72975 & Known & 13.45 & -1.515 & 1.1979 & 48.78 & 1.541 & $2290$ & $^{+109}_{-97}$ \\
        J181752$-$580749 & 274.4647083 & -58.13033333 & Known & 12.45 & -1.52 & 2.07 & 1.32 & 0.927 & $2656$ & $^{+135}_{-110}$ \\
        J005255$+$633515 & 13.22923194 & 63.58755732 & Pipeline & 13.60 & -0.6458 & 1.39221773 & 26.55 & 1.229 & $2180$ & $^{+51}_{-57}$ \\
        J062510$-$075341 & 96.29314956 & -7.89464499 & Pipeline & 13.678 & 0.1687 & 1.173 & 23.68 & 1.444 & $1904$ & $^{+66}_{-52}$ \\
        J124745$-$622756 & 191.9366145 & -62.4654549 & Pipeline & 13.543 & -4.633 & 0.397 & 54.89 & 1.143 & $3866$ & $^{+202}_{-246}$ \\
        J160757$-$574540 & 241.9885247 & -57.76106277 & Pipeline & 13.902 & -0.098 & 1.083 & 36.67 & 0.858 & $2557$ & $^{+80}_{-75}$ \\
        J175912$-$120956 & 269.7994218 & -12.16566391 & Pipeline & 13.474 & 0.54444 & 0.488 & 204.92 & 0.946 & $1386$ & $^{+35}_{-26}$ \\
        \addlinespace
        \multicolumn{10}{l}{\textbf{Dippers}} \\
        J070519$+$061219 & 106.328990 & 6.205306 & Known & 13.73 & 2.65 & 0.559 &  & 0.971 & $1010$ & $^{+15}_{-23}$ \\
        J081523$-$385923 & 123.8475 & -38.98980556 & Known & 13.94 & 3.773 & 0.36 &  & 1.104 & $556$ & $^{+3}_{-3}$ \\
        J085816$-$430955 & 134.564850 & -43.165320 & Known & 12.31 & -0.601 & -0.18 & 24.85 & 5.440 & $1031$ & $^{+73}_{-68}$ \\
        J114712$-$621057 & 176.7999167 & -62.17702778 & Known & 13.04 & -0.186 & -0.05 &  & 0.903 & $2607$ & $^{+103}_{-85}$\\
        J181707$-$164819 & 274.27809 & -16.80539 & Known & 13.6192 & 0.9869 & 0.1424 &  & 1.02 & $1515$ & $^{+39}_{-38}$ \\
        J183153$-$284827 & 277.96999 & -28.80754 & Known & 13.6566 & 0.52797 & 1.105 & 2.85 & 0.844 & $2002$ & $^{+50}_{-68}$ \\
        J184916$-$473251 & 282.31475 & -47.54741667 & Known & 13.23 & -2.2339 & 1.65 & 6.29 & 1.597 & $6778$ & $^{+541}_{-541}$ \\
        J042214$+$152530 & 65.55736276 & 15.42493885 & Pipeline & 13.346 & 3.466 & 1.170 & 54.18 & 1.305 & $288$ & $^{+2}_{-2}$ \\
        J080327$-$261620 & 120.8622157 & -26.27215412 & Pipeline & 13.315 & -0.67 & 0.016 &  & 0.977 & $4760$ & $^{+300}_{-328}$ \\
        J174328$+$343315 & 265.8684124 & 34.55405423 & Pipeline & 13.369 & 1.2388 & 0.946 & 84.84 & 1.087 & $1904$ & $^{+46}_{-42}$ \\
        J202402$+$383938 & 306.0094122 & 38.66055087 & Pipeline & 13.114 & -0.81 & 0.883 &  & 1.459 & $1827$ & $^{+52}_{-46}$ \\
        \bottomrule
    \end{tabular}
    \label{table:target_param}
\end{table*}

\begin{table*}
    \centering
    \caption{The objects are sorted as in Table \ref{table:target_param}. Each target has a source id, the right ascension, the declination, the number of dips recovered by our pipeline, the maximum dip duration, the minimum period of the dips the for the single eclipse binaries, start date of the most recent dip, and the maximum dip depth.}
    \sisetup{table-auto-round,
     group-digits=false, table-number-alignment = center, separate-uncertainty = true}
    \setlength{\tabcolsep}{8pt}
    \begin{tabular}{l S[table-format=3.5] S[table-format=3.5] l S[table-format=2.0] c S[table-format=4.1] l S[table-format=2.2]}
        \toprule
        {Source} & 
        {RA} & 
        {DEC} & 
        \multicolumn{1}{c}{\shortstack{{Search} \\ {Method}}} &
        \multicolumn{1}{c}{\shortstack{{Number of} \\ {Dips}}} &  
        \multicolumn{1}{c}{\shortstack{{Maximum} \\ {Dip Duration}}} & 
        \multicolumn{1}{c}{\shortstack{{Minimum} \\ {Period}}} &
        \multicolumn{1}{c}{\shortstack{{Start of Most} \\ {Recent Dip}}} &  
        \multicolumn{1}{c}{\shortstack{{Maximum} \\ {Depth}}} \\
        {} & {[deg]} & {[deg]} & {} & {} & {[days]} & {[days]} & {} & {[mag]} \\
        \midrule
        \multicolumn{8}{l}{\textbf{Single Eclipse Binaries}} \\
        J073234$-$200049 & 113.142375 & -20.01358333 & Known & 1 & 40.9 & 3012 & 07/10/2023 & 1.375 \\
        J175602$+$013135 & 269.0079973 & 1.52628966 & Known & 1 & 94.8 & 1949 & 22/02/2024 & 2.28201 \\
        J223332$+$565552 & 338.384125 & 56.93108333 & Known & 1 & 19.9 & 1694 & 15/11/2023 & 0.65241 \\
        J073924$-$272916 & 114.8503576 & -27.48776143 & Pipeline & 1 & 26.9 & 1307 & 25/02/2023 & 0.33792 \\
        J074007$-$161608 & 115.0276126 & -16.26896725 & Pipeline & 1 & 4.0 & 769 & 21/09/2022 & 0.3275 \\
        J094848$-$545959 & 147.1987289 & -54.99963893 & Pipeline & 1 & 7.7 & 1165 & 20/03/2019 & 0.31774 \\
        J162209$-$444247 & 245.5357562 & -44.71308061 & Pipeline & 1 & 3.1 & 544 & 25/04/2023 & 0.60564 \\
        J183210$-$173432 & 278.0424857 & -17.57545433 & Pipeline & 1 & 41.0 & 1516 & 02/09/2018 & 0.30911 \\
        J183606$-$314826 & 279.0252466 & -31.80714478 & Pipeline & 1 & 10.7 & 753 & 10/03/2022 & 0.38631 \\
        J190316$-$195739 & 285.8184954 & -19.9609327 & Pipeline & 1 & 15.3 & 1992 & 19/08/2019 & 0.39416 \\ 
        J205245$-$713514 & 313.1890422 & -71.58708665 & Pipeline & 1 & 2.0 & 756 & 14/04/2023 & 0.3801 \\
        J212132$+$480140 & 320.3840472 & 48.02785275 & Pipeline & 1 & 57.8 & 1348 & 09/09/2021 & 0.37429 \\
        J225702$+$562312 & 344.2585271 & 56.38663345 & Pipeline & 1 & 6.0 & 1002 & 08/06/2019 & 0.39525 \\
        \addlinespace
        \multicolumn{8}{l}{\textbf{Multiple Eclipse Binaries}} \\
        J005437$+$644347 & 13.652625 & 64.72975 & Known & 5 & 10.9 & & 01/12/2024 & 0.58511 \\
        J181752$-$580749 & 274.4647083 & -58.13033333 & Known & 1 & 86.8 & & 07/08/2024 & 0.6334 \\
        J005355$+$633515 & 13.22923194 & 63.58755732 & Pipeline & 3 & 10.1 & & 06/08/2020 & 0.48852 \\
        J062510$-$075341 & 96.29314956 & -7.89464499 & Pipeline & 1 & 4.3 & & 14/08/2019 & 0.6737 \\
        J124745$-$622756 & 191.9366145 & -62.4654549 & Pipeline & 2 & 7.2 & & 14/05/2024 & 0.37694 \\
        J160757$-$574540 & 241.9885247 & -57.76106277 & Pipeline & 2 & 8.3 & & 17/04/2024 & 0.30637 \\
        J175912$-$120956 & 269.7994218 & -12.16566391 & Pipeline & 2 & 0.0 & & 12/05/2022 & 0.477 \\
        \addlinespace
        \multicolumn{8}{l}{\textbf{Dippers}} \\
        J070519$+$061219 & 106.328990 & 6.205306 & Known & 16 & 274.8 & & 23/08/2024 & 3.8379698 \\
        J081523$-$385923 & 123.8475 & -38.98980556 & Known & 5 & 608.1 & & 30/03/2021 & 2.03064 \\
        J085816$-$430955 & 134.564850 & -43.165320 & Known & 1 & > 200 days, Ongoing & & 06/04/2025 & 0.98509 \\
        J114712$-$621057 & 176.7999167 & -62.17702778 & Known & 0 & 1282.8 & & 20/11/2020 & 0.32485\\
        J181707$-$164819 & 274.27809 & -16.80539 & Known & 5 & > 200 days, Ongoing & & 11/05/2025 & 1.437668 \\
        J183153$-$284827 & 277.96999 & -28.80754 & Known & 6 & > 200 days, Ongoing & & 06/03/2025 & 2.53 \\
        J184916$-$473251 & 282.31475 & -47.54741667 & Known & 1 & 273.1 & & 08/07/2021 & 0.32731 \\
        J042214$+$152530 & 65.55736276 & 15.42493885 & Pipeline & 6 & 44.2 & & 04/10/2022 & 1.02054 \\
        J080327$-$261620 & 120.8622157 & -26.27215412 & Pipeline & 11 & > 1400 days, Ongoing & & 31/01/2022 & 0.48963 \\
        J174328$+$343315 & 265.8684124 & 34.55405423 & Pipeline & 4 & 22.1 & & 25/02/2020 & 0.323697 \\
        J202402$+$383938 & 306.0094122 & 38.66055087 & Pipeline & 7 & 1709.6 & & 15/04/2018 & 0.75821 \\
        \bottomrule
    \end{tabular}
    \label{table:dip_param}
\end{table*}

\section*{Acknowledgements}

We thank the anonymous reviewer for their comments. DMR acknowledges support from the OSU Presidential Fellowship. CSK, KZS, and DMR are supported by NSF grants AST-2307385 and 2407206. B.J.S is supported with funds from the NSF (grant AST-2407205). S.D. acknowledges the National Natural Science Foundation of China (Grant No. 12133005) and the New Cornerstone Science Foundation through the XPLORER PRIZE.


We thank Las Cumbres Observatory and its staff for their continued support of ASAS-SN. ASAS-SN is funded by Gordon and Betty Moore Foundation grants GBMF5490 and GBMF10501 and the Alfred P. Sloan Foundation grant G-2021-14192. 

This work presents results from the European Space Agency space mission Gaia. Gaia data are being processed by the Gaia Data Processing and Analysis Consortium (DPAC). Funding for the DPAC is provided by national institutions, in particular the institutions participating in the Gaia MultiLateral Agreement.

This work included the use of the VizieR catalogue access tool, CDS, Strasbourg, France \citep{Ochsenbein_2000}.

\clearpage
\bibliographystyle{mnras}
\bibliography{main} 

\end{document}